\documentclass[reprint,amsmath,amssymb,aps,prb,superscriptaddress,longbibliography]{revtex4-1}
\usepackage{array}
\usepackage{bm}
\usepackage{color}
\usepackage{float}
\usepackage{graphicx}
\usepackage{hyperref}
\hypersetup{colorlinks=true,allcolors=blue}
\usepackage{natbib}
\usepackage{newtxtext}
\usepackage{newtxmath}
\usepackage[caption=false]{subfig}
\begin{document}

%Title of paper
\title{Magnetic vortex liquid and anomalous thermal Hall conductivity \\ in frustrated magnets with bond-dependent interactions}

\author{Li Ern Chern}
\affiliation{Department of Physics, University of Toronto, Toronto, Ontario M5S 1A7, Canada}

\author{Finn Lasse Buessen}
\affiliation{Department of Physics, University of Toronto, Toronto, Ontario M5S 1A7, Canada}

\author{Yong Baek Kim}
\affiliation{Department of Physics, University of Toronto, Toronto, Ontario M5S 1A7, Canada}
%\email[]{}
%\thanks{}
%\altaffiliation{}
	
%Collaboration name if desired (requires use of superscriptaddress
%option in \documentclass). \noaffiliation is required (may also be
%used with the \author command).
%\collaboration can be followed by \email, \homepage, \thanks as well.
%\collaboration{}
%\noaffiliation

%\date{\today}

\begin{abstract}
Recently, the observation of large thermal Hall conductivities in correlated insulators with no apparent broken symmetry have generated immense interest and debates on the underlying ground states. Here, considering frustrated magnets with bond-dependent interactions, which are realized in the so-called Kitaev materials, we theoretically demonstrate that a large thermal Hall conductivity can originate from a classical ground state without any magnetic order. We discover a novel liquid state of magnetic vortices, which are inhomogeneous spin textures embedded in the background of polarized spins, under out-of-plane magnetic fields. In the classical regime, different configurations of vortices form a degenerate manifold. We study the static and dynamical properties of the magnetic vortex liquid state at zero and finite temperatures. In particular, we show that the spin excitation spectrum resembles a continuum of nearly flat Chern bands, which ultimately leads to a large thermal Hall conductivity. Possible connections to experiments are discussed.
\end{abstract}

% insert suggested PACS numbers in braces on next line
% insert suggested keywords - APS authors don't need to do this
%\keywords{}

%\maketitle must follow title, authors, abstract, \pacs, and \keywords
\maketitle

% body of paper here - Use proper section commands
% References should be done using the \cite, \ref, and \label commands

\section{\label{introduction}Introduction}
Thermal Hall conductivity in correlated insulators is an important diagnostic tool to uncover the underlying many-body ground states. Recent observations of the half-quantized thermal Hall conductivity in the spin liquid candidate $\alpha$-RuCl$_3$\cite{s41586-018-0274-0,2001.01899}, and a large thermal Hall signal in cuprates\cite{s41586-019-1375-0}, are just two prominent examples. As the charge transport is absent in these systems, collective degrees of freedom such as magnons and phonons, or charge-neutral fractionalized excitations in quantum spin liquids, are most likely to be the sources of such effects\cite{KITAEV20062,PhysRevX.8.031032,PhysRevLett.121.147201,PhysRevResearch.2.033180}. In this work, we present a novel route to obtain a large thermal Hall conductivity in frustrated magnets with nearest neighbor bond-dependent spin interactions. Such interactions are realized in the so-called Kitaev materials\cite{PhysRevLett.102.017205,PhysRevLett.112.077204,s42254-019-0038-2,Janssen_2019}, which includes $\alpha$-RuCl$_3$\cite{PhysRevB.90.041112,PhysRevB.91.144420} and Na$_2$IrO$_3$\cite{PhysRevLett.105.027204,Katukuri_2014,nphys3322}. In particular, we investigate the $K \Gamma$ model on the honeycomb lattice with an external magnetic field along the $[111]$ direction, where $K>0$ is the antiferromagnetic Kitaev interaction\cite{s41467-019-08459-9,PhysRevB.100.144445,1809.08247} and $\Gamma<0$ is the symmetric anisotropic interaction. As discussed below, a more realistic generalization of this model is also studied. We consider the classical limit of these models, which enables numerical simulations on large systems. \\

At zero temperature, we discover a degenerate manifold of distinct spin configurations of magnetic vortices in an appreciable window of magnetic fields, right below the fully polarized state. In each configuration, the vortices are embedded in the background of polarized spins, and they do not form a crystalline order. We call this manifold of states \textit{magnetic vortex liquid} as they form a thermal ensemble at finite temperatures. The vortices decrease in number as the field increases, eventually disappear completely as the system enters the fully polarized state. \\

The magnons in the background polarized state sees the vortices as a source of fictitious magnetic flux via the Berry phase effect\cite{rspa.1984.0023,PhysRevB.72.024456,PhysRevB.87.024402,PhysRevB.91.104435,TATARA2019208,RevModPhys.82.1959,1.4870695}. In order to understand the spin excitation spectrum, we first use a mean-field approximation, where the magnetic flux carried by the vortices is spread over the entire system. In other words, the magnons see an averaged, uniform magnetic flux. The resulting magnon bands consist of many relatively flat sub-bands and form a dense spin excitation spectrum, which looks very much like a continuum. It is found that the magnon bands are topologically nontrivial - they carry finite Chern numbers, which contribute to a large thermal Hall conductivity\cite{PhysRevLett.104.066403,1.4959815,PhysRevB.94.094405,Owerre_2017,cjp-2018-0059}. \\

At finite temperatures, we employ Monte Carlo simulations and the Landau-Lifshitz equations of motion to investigate the static and dynamical properties of the magnetic vortex liquid. The vortex liquid is found to persist at finite temperatures, and it is separated by a crossover, rather than a sharp transition, from the high temperature paramagnetic state. The equal-time spin correlator indeed reveals the characteristics of a liquid state. On the other hand, the dynamical structure factor exhibits a continuum of spin excitations, which strongly resembles the excitation spectrum obtained in the mean-field approximation. \\

Furthermore, we investigate the relevance of the magnetic vortex liquid to $\alpha$-RuCl$_3$ and similar materials. Ref.~\onlinecite{PhysRevB.92.024413} identifies a self-dual transformation of the $J K \Gamma \Gamma'$ model, where $J$ is the Heisenberg interaction and $\Gamma'$ is another off diagonal interaction. Under this transformation, the $J K \Gamma \Gamma'$ model is mapped to the same model yet with a different set of parameters $\tilde{J}$, $\tilde{K}$, $\tilde{\Gamma}$, and $\tilde{\Gamma}'$. We verify that in the dual model of the $K \Gamma$ model with $K>0$ and $\Gamma<0$, the sign of $\tilde{K}$ is always negative (i.e.~ferromagnetic), while that of $\tilde{\Gamma}$ is positive for $\lvert \Gamma / K \rvert < 4/5$. Such a sign structure is the same as that in $\alpha$-RuCl$_3$\cite{PhysRevB.93.155143,PhysRevB.93.214431,PhysRevB.96.115103,s41467-017-01177-0,s41567-020-0874-0}. The resulting ground states of the dual models in a $[111]$ magnetic field are in one-to-one correspondence. More interestingly, if we continuously evolve the dual model towards the parametric regime relevant to $\alpha$-RuCl$_3$, we find a significant extent of the vortex liquid at high fields, while the ground state at zero and low fields becomes the zigzag magnetic order\cite{PhysRevB.91.144420,PhysRevB.92.235119}. \\

\begin{figure*}
\includegraphics[scale=0.36]{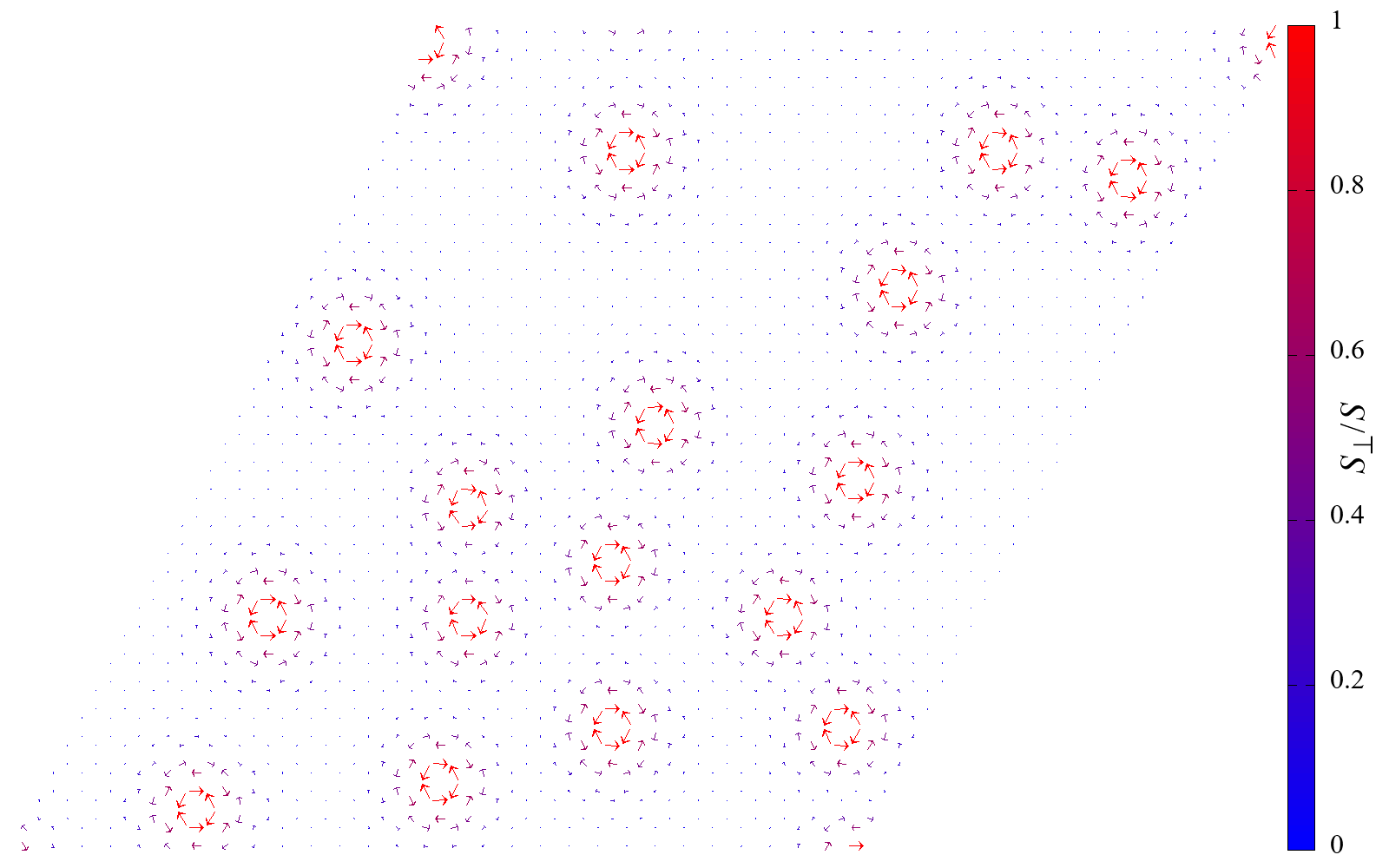}
\caption{\label{phi020h112}A sample spin configuration of the magnetic vortex liquid obtained via classical simulated annealing, at $\varphi/\pi=0.2$ and $h/S=1.12$, on a lattice of $30 \times 30 \times 2$ sites with periodic boundary conditions. The $[111]$ direction points out of plane, along which most spins are polarized. Color of the spins indicates the magnitude of the (normalized) transverse magnetization $S_\perp/S$. More figures can be found in Supplemental Material.}
\end{figure*}

The rest of the paper is organized as follows. In Sec.~\ref{model}, we introduce the $K \Gamma$ model in a $[111]$ magnetic field. In Sec.~\ref{results}, we present the main results of our study. We discuss the emergence of magnetic vortex liquid and its physical properties, which include the real space spin configuration, the equal time spin correlator, and the dynamical spin structure factor. We also demonstrate, through a mean-field analysis, how a large thermal Hall conductivity can arise from the vortex liquid. In Sec.~\ref{methods}, we provide technical justifications to the mean-field analysis as well as other methodological details. In Sec.~\ref{materialrelevance}, we investigate the extent of vortex liquid towards realistic models. In Sec.~\ref{discussion}, we comment on the connections between our work and existing experimental and theoretical studies.

\section{\label{model}Model}
We start with the classical $K \Gamma$ model on the honeycomb lattice, with $K>0$ and $\Gamma<0$, under an external magnetic field $\mathbf{h}$ along the $[111]$ direction (perpendicular to the lattice plane),
\begin{equation} \label{KGhspinmodel}
H = \sum_\lambda \sum_{\langle ij \rangle \in \lambda} \left[ K S_i^\lambda S_j^\lambda + \Gamma \left( S_i^\mu S_j^\nu + S_i^\nu S_j^\mu \right) \right] - \sum_i \mathbf{h} \cdot \mathbf{S}_i ,
\end{equation}
where $(\lambda,\mu,\nu)$ is a cyclic permutation of $(x,y,z)$, $K$ is the Kitaev interaction, $\Gamma$ is the symmetric anisotropic interaction, and the field $\mathbf{h}=h(1,1,1)/\sqrt{3}$. The spins $\mathbf{S}_i$ are treated as three-component vectors of fixed magnitude, $\lvert \mathbf{S}_i \rvert = S$. We can specify $S=1/2$ for example, but we will keep the factor $S$ explicit when discussing quantities such as field, energy, and temperature later. We adopt the trigonometric parametrization of the interactions, $K=\cos \varphi, \Gamma=-\sin \varphi$, with $\varphi \in [0,\pi/2]$.

\section{\label{results}Results}

\subsection{\label{emergentvortexliquid}Emergence of Magnetic Vortex Liquid}
We use classical simulated annealing to obtain the zero temperature spin configuration on a lattice of $L \times L$ unit cells (or $L \times L \times 2$ sites), up to $L=36$. The ground state phase diagram is explored as a function of $\phi$ and $h$. Periodic boundary conditions are imposed to reduce finite size effect. We refer interested readers to Ref.~\onlinecite{PhysRevLett.117.277202,PhysRevResearch.2.013014} for details of the simulated annealing calculation. \\

Away from the antiferromagnetic Kitaev limit and the ferromagnetic $\Gamma$ limit, i.e.~for $\varphi \neq 0, \pi/2$, we observe a rather unusual phase at high fields right below the polarized state. This phase is characterized by \textit{magnetic vortices}, each consisting of a cluster of spins, forming an irregular, fluid-like pattern as opposed to a crystalline order. The spin configuration with these vortices is a \textit{textured ferromagnet}\cite{PhysRevB.72.024456}, meaning that most of the spins (apart from the cores of the vortices) are nearly or fully polarized. We call such a phase \textit{magnetic vortex liquid}. An example of the real space spin configuration is shown in Fig.~\ref{phi020h112}. The density of vortices in the lattice depends on the magnitude of the field. At high fields near the polarized regime, the vortices are dilute (i.e.~well separated from each other), and there is essentially no interaction between them (e.g. see Fig.~S1a in Supplemental Material\cite{SM}). These non-interacting vortices usually appear as low energy excitations of the fully polarized state. However, as the field is lowered, the vortices become dense enough such that their boundaries overlap (e.g. see Fig.~S1b in Ref.~\onlinecite{SM}). The spin configuration formed by these interacting vortices has lower energy than that of the fully polarized state. \\

\begin{figure}
\includegraphics[scale=0.25]{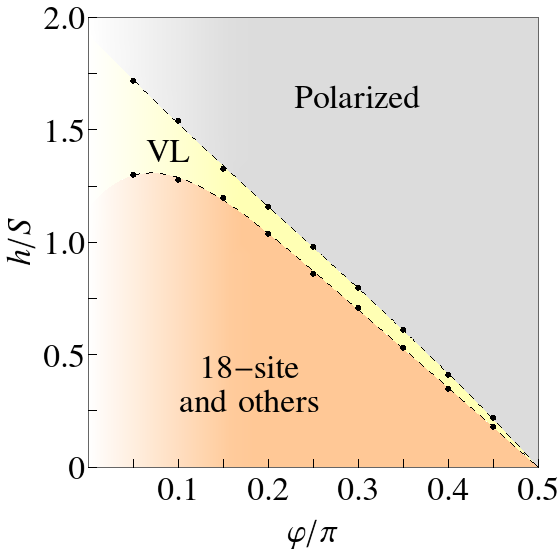}
\caption{\label{phasediagram}The magnetic vortex liquid appears in the parametric regime between the dashed lines. The phase boundaries are by no means exact. In particular, the transition from the vortex liquid to the polarized state is most likely a crossover. At low and intermediate fields, we find a relatively dominant 18-site order as well as other phases.}
\end{figure}

Fig.~\ref{phasediagram} is a schematic phase diagram which indicates the range of parameters where the vortex liquid, labeled as ``VL'', appears. Certainly, there exist other phases in the parametric region below the vortex liquid. Among them is a dominant 18-site order (i.e.~a magnetic order with 18 sublattices per unit cell), a vortex crystal (more discussions later when we present the results at finite temperatures), and possibly some incommensurate order. However, these phases are not our main concern in this work, so we label them collectively as ``18-site and others''. The phase boundaries of the vortex liquid, indicated by the dashed lines in Fig.~\ref{phasediagram}, are not meant to be exact, due to the following reasons. As the field increases, the density of vortices decreases continuously to zero, and the system gradually becomes polarized without encountering a first order transition. As the field decreases, the precise occurrence of phase transition(s) to other nontrivial phase(s), e.g.~an incommensurate order, is obscured by the large size of the vortex and the finiteness of the system. \\

Some remarks are in order. First, the emergence of magnetic vortices requires the presence of both $K$ and $\Gamma$ interactions. In the Kitaev limit $\varphi=0$, the system appears to be disordered all the way from $h/S=0$ to $2 K$, at which it achieves full polarization. Within this range of fields, the longitudinal magnetization is a linear function of the field, indicating a paramagnetic behavior. In the $\Gamma$ limit $\varphi=\pi/2$, the ground state at $h=0$ is a degenerate manifold that includes the ferromagnet with spins aligned in the $[111]$ direction, which is certainly selected when $h \neq 0$. The vortices start to appear at $\varphi$ as small as $0.05 \pi$, and persist until $\varphi$ as large as $0.45 \pi$. Second, the size of a single vortex decreases, and its structure simplifies, as $\varphi$ or $\lvert \Gamma/K \rvert$ increases. It can be as large as the system size (e.g. see Fig.~S2a in Ref.~\onlinecite{SM}) or a cluster of about $24$ spins (e.g. see Fig.~S2b in Ref.~\onlinecite{SM}). The spins are mostly lying in plane at the core of the vortex, but they gradually align with the field as we move further away from the core. Third, together with the critical field to full polarization, the fields at which the vortex liquid appears decrease with increasing $\varphi$. The window of field within which the vortex liquid appears also shrinks in width with increasing $\varphi$.

\subsection{\label{meanfieldtheory}Mean-Field Theory}
One may ask what physical properties can be extracted from the vortex liquid. To allow analytical progress, some approximations have to be made. As mentioned previously, the magnetic vortex liquid is a essentially a textured ferromagnet - the vortices are inhomogeneous spin textures embedded in the background of polarized spins. In other words, most of the spins in the vortex liquid are nearly or completely aligned with the field, which encourages us to treat the system as the polarized state on average. However, to observe the nontrivial physical effects due to the presence of magnetic vortices, their nonuniform spin textures should somehow be taken into account, at least at the mean-field level. Below we outline our program in a concise manner, while relaying the details of derivations and calculations to Sec.~\ref{meanfieldmethod}. \\

\begin{figure}
\includegraphics[scale=0.36]{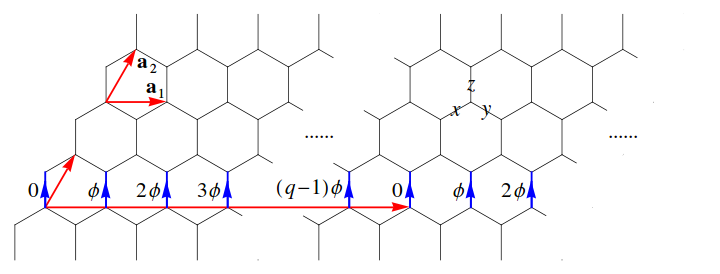}
\caption{\label{optimalgauge}The optimal gauge is a choice of the $U(1)$ gauge field which produces a uniform magnetic flux. Here, the labels $x$, $y$, and $z$ represent the three bond types in the $K \Gamma$ model. The magnons only gain an additional $U(1)$ phase of $n_1 \phi$ when hopping along the $z$ bonds, as indicated by the blue arrows. Such a gauge choice breaks the translational symmetry of the original lattice (with the primitive vectors $\mathbf{a}_1$ and $\mathbf{a}_2$). The resulting magnetic unit cell contains $2q$ sublattices for a flux of $\phi=- 2\pi p/q$ per unit hexagon, where $p$ and $q$ are relatively prime. With the mean-field approximation, we have mapped the vortex liquid to the bosonic Hofstadter problem.}
\end{figure}

We first start from the polarized state, in which all spins are aligned in the $[111]$ direction, and then incorporate the effect of magnetic vortices in the linear spin wave theory\cite{PhysRev.58.1098,Jones_1987}, as follows. In general, a smoothly varying spin texture gives rise to a real space Berry phase for magnons hopping on the lattice\cite{PhysRevB.72.024456,PhysRevB.87.024402,PhysRevB.91.104435,TATARA2019208}. While magnons, unlike electrons and other charged particles, are not coupled to the vector potential of the external magnetic field, they can experience a fictitious gauge field due to the aforementioned Berry phase effect. We elucidate this idea with a judicious choice of the parametrization $\varphi=\tan^{-1} (1/2)$ or $K=-2 \Gamma$, where the derivation of the effective gauge field greatly simplifies. At such parametrization, all magnon pairings vanishes and the linear spin wave Hamiltonian of the background polarized state reduces to a tight binding model,
\begin{equation}\label{bosontightbinding}
\frac{H}{S} = - \Gamma \sum_{\langle ij \rangle} \left( b_i^\dagger b_j + b_j^\dagger b_i \right) + h \sum_i b_i^\dagger b_i .
\end{equation}
Next, we show from the continuum model that the spatial variation of spins in the vortex liquid gives rise to a $U(1)$ gauge field of the form $A_\mu (\mathbf{r}) = - \cos \theta_\mathbf{r} \partial_\mu \phi_\mathbf{r}$ which couples to the magnons\cite{TATARA2019208}, similar to how the vector potential couples to the electrons in quantum electrodynamics. Here $\mu=x,y$, while $\theta_\mathbf{r}$ and $\phi_\mathbf{r}$ are the angles parametrizing the spin orientation at position $\mathbf{r}$. In other words, magnons hopping on the lattice see a fictitious magnetic flux originating from the nonuniform spin textures in the vortex liquid\cite{PhysRevB.72.024456,PhysRevB.87.024402,PhysRevB.91.104435}. We calculate the total flux of the vortex liquid and spread it uniformly over the lattice. Each vortex carries a flux of $- 4 \pi$, and since the number of vortices and the number of unit hexagons are integers, the flux per unit hexagon is equal to a rational number times the flux quantum, i.e.~$\phi=- 2 \pi p/q$ where $p$ and $q$ are relatively prime. \\

\begin{figure}
\includegraphics[scale=0.25]{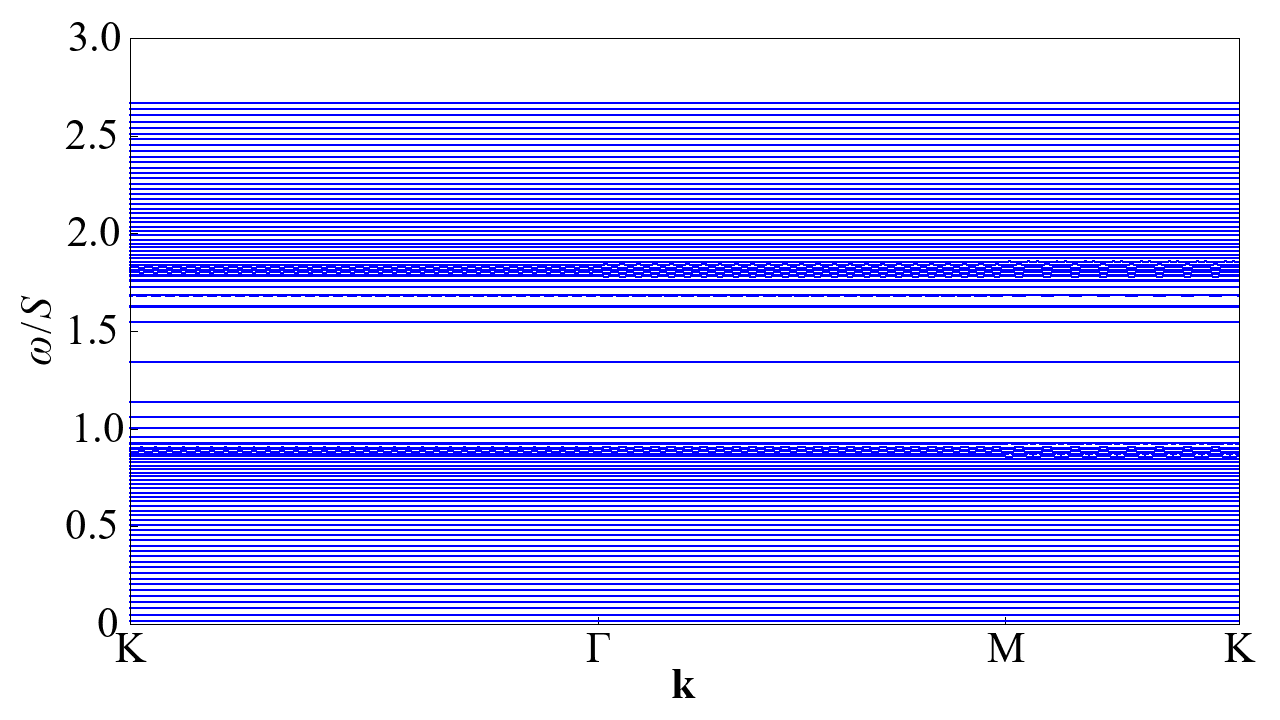}
\caption{\label{dispersion}The magnon energy spectrum along high symmetry directions in the first Brillouin zone, for the vortex liquid at $\varphi = \tan^{-1}(1/2)$ and $h/S = 1.3$. Since the flux per unit hexagon is $\phi= -2 \pi \times 1/50$, there are in total $2 \times 50=100$ bands. The bands are nearly flat. Dispersive features are only visible in few bands at intermediate energies.}
\end{figure}

It is the $U(1)$ gauge field, instead of the fictitious magnetic flux, which couples directly to the magnons hopping on the lattice. Therefore, after making the uniform flux approximation, we have to choose a gauge field $\mathbf{A}$ such that the curl $\nabla \times \mathbf{A}$ produces the uniform flux. To this end, we use the optimal gauge\cite{PhysRevB.74.155415,PhysRevB.86.235411} in which the magnon gains a $U(1)$ phase (via Peierls substitution\cite{Peierls_1933,PhysRev.84.814,PhysRevB.14.2239,PhysRevLett.63.907}) of $n_1 \phi$ only when hopping along a $z$ bond, where $n_1$ is an integer labeling a site $\mathbf{R}$ on the underlying Bravais lattice of the honeycomb lattice along the $\mathbf{a}_1$ direction, i.e.~$\mathbf{R} = n_1 \mathbf{a}_1 + n_2 \mathbf{a}_2$. \eqref{bosontightbinding} becomes
\begin{equation}\label{hofstadtermagnon}
\frac{H}{S} = - \sum_{\langle ij \rangle} \left( \Gamma_{ij} b_i^\dagger b_j + \Gamma_{ij}^* b_j^\dagger b_i \right) + h \sum_i b_i^\dagger b_i ,
\end{equation}
with $i \in 0$, $j \in 1$, $\Gamma_{ij}=\Gamma \exp (-i n_1 \phi)$ for $\langle ij \rangle \in z$, and $\Gamma_{ij}=\Gamma$ for $\langle ij \rangle \in x,y$. The lattice curl of the $U(1)$ gauge field for any unit hexagon is $(n_1+1 - n_1) \phi = \phi$, which is exactly what we want. The optimal gauge is illustrated in Fig.~\ref{optimalgauge}. \eqref{hofstadtermagnon} is essentially the bosonic analogue of Hofstadter problem\cite{PhysRevB.14.2239,PhysRevB.95.125429,OWERRE201893}, which is no longer the problem of a trivial polarized state. If we plot the allowed energy levels as a function of $\phi$, we will obtain the Hofstadter butterfly. \\

Finally, we stress that $\varphi=\tan^{-1} (1/2)$ is just a convenient parametric choice for various calculations, while the real space Berry phase generically arises whenever we have an inhomogeneous spin texture\cite{PhysRevB.72.024456,PhysRevB.87.024402,PhysRevB.91.104435,TATARA2019208}, e.g.~in the vortex liquids at other $\varphi$.

\subsection{\label{thermalhalleffect}Thermal Hall Effect}
We put the above ideas into action by applying them to a specific example. We consider a sample spin configuration of the vortex liquid at $\varphi=\tan^{-1} (1/2)$ and $h/S=1.3$, which has about $9$ magnetic vortices on a lattice of size $L=30$ and a total flux of $-4 \pi \times 9$ (see Fig.~S3 in Ref.~\onlinecite{SM}). Averaging the flux over the lattice, the flux per unit hexagon is $\phi=-4 \pi \times 9/30^2 = -2 \pi \times 1/50$. Therefore, we have $p=1$ and $q=50$. \\

The optimal gauge breaks the translational symmetry by elongating the magnetic unit cell in the $\mathbf{a}_1$ direction. Since each magnetic unit cell contains $2q$ sites, there are in total $2q$ bands in the reciprocal space. Fig.~\ref{dispersion} shows the magnon spectrum along a high symmetry cut in the first Brillouin zone. One immediately sees that the bands are nearly flat - the variation of energy within any band is negligible when compared to the intrinsic energy scale $\sqrt{K^2+\Gamma^2}=1$. \\
%This will simplify the calculation of the thermal Hall conductivity, as discussed below. \\

\begin{figure}
\includegraphics[scale=0.36]{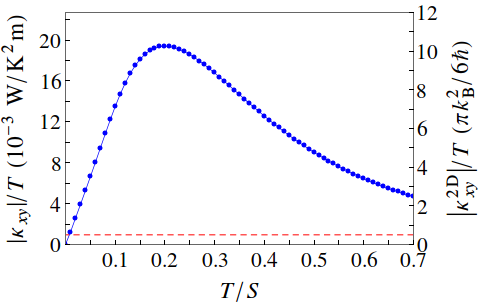}
\caption{\label{thermalhalldata} The thermal Hall conductivity (in absolute value) due to magnons in the vortex liquid at $\varphi = \tan^{-1}(1/2)$ and $h/S = 1.3$, as a function of temperature (joined blue dots). The half-quantized thermal hall conductance $\kappa_{xy}^\mathrm{2D}/T=(1/2)(\pi/6)(k_\mathrm{B}^2/\hbar)$ is also indicated (red dashed line) for comparison.}
\end{figure}

Once we obtain the magnon spectrum, we can calculate the thermal Hall conductivity, which plays an important role in probing the many-body ground state of strongly correlated systems and bridging theoretical models to realistic materials. The thermal Hall conductivity due to magnons is given by\cite{PhysRevLett.106.197202,PhysRevB.89.054420,JPSJ.86.011010}
\begin{equation} \label{thermalhallconductivityformula}
\begin{aligned}[b]
\kappa_{xy} &= -\frac{k_\mathrm{B}^2 T}{\hbar V} \sum_{n\mathbf{k}} \left \lbrace c_2 \left[ g \left( \varepsilon_{n\mathbf{k}} \right) \right] - \frac{\pi^2}{3} \right \rbrace \Omega_{n \mathbf{k}} \\
&\approx -\frac{1}{2 \pi} \frac{k_\mathrm{B}^2 T}{\hbar l_c} \sum_{n} c_2 \left[ g \left( \varepsilon_{n\mathbf{0}} \right) \right] \left[ \frac{1}{2 \pi} \sum_\mathbf{k} \frac{(2 \pi)^2}{A} \Omega_{n \mathbf{k}} \right] \\
&= -\frac{1}{2 \pi} \frac{k_\mathrm{B}^2 T}{\hbar l_c} \sum_{n} c_2 \left[ g \left( \varepsilon_{n\mathbf{0}} \right) \right] C_n ,
\end{aligned}
\end{equation}
where the $V$ is the total volume of the system, i.e. the total area $A$ times the interlayer distance $l_c$, which we set to be $5 \textrm{\AA}$ (c.f. $l_c=5.72 \textrm{\AA}$ in $\alpha$-RuCl$_3$), $c_2 (x) = (1+x) \lbrace \ln [(1+x) / x] \rbrace ^2 - (\ln x)^2 - 2 \mathrm{Li}_2 (-x)$, $g (\varepsilon) = [\exp(\varepsilon/k_\mathrm{B} T) - 1]^{-1}$ is the Bose-Einstein distribution, $\Omega_{n \mathbf{k}}$ is the reciprocal space Berry curvature of the $n$th band at momentum $\mathbf{k}$, and $C_n$ is the Chern number of the $n$th band. In the second equality, we have dropped the term $-\pi^2/3$, due to the fact that the summation of all Chern numbers is zero\cite{PhysRevB.87.174427}. Furthermore, we have approximated $\varepsilon_{n \mathbf{k}}$ to be constant for a given $n$, and take the energy at $\mathbf{k}=\mathbf{0}$ as a representative, thus removing the $\mathbf{k}$ dependence. This is valid when the temperature scale is much larger than the variation of energy within the thermally populated bands, a condition which always holds in the magnon spectrum under study (see Fig.~\ref{dispersion}). Lastly, we introduce the thermal Hall conductance $\kappa_{xy}^\mathrm{2D}=\kappa_{xy} l_c$, which is independent of the interlayer distance. \\

When the field is along the $[111]$ direction, the resulting thermal Hall conductivity is negative. For convenience of comparison to the half-quantized thermal Hall conductance $\kappa_{xy}^\mathrm{2D}/T=(1/2)(\pi/6)(k_\mathrm{B}^2/\hbar)$ due to Majorana fermions\cite{KITAEV20062,s41586-018-0274-0,2001.01899,PhysRevLett.120.217205}, which is positive, we plot the absolute value $\lvert \kappa_{xy} \rvert/T$ as a function of $T$, for the vortex liquid at $\varphi=\tan^{-1} (1/2)$ and $h/S=1.3$, in Fig.~\ref{thermalhalldata}. To obtain a positive thermal Hall conductivity, one can reverse the field direction (see Sec.~\ref{chernnumber}). \\

The magnon thermal Hall conductivity is huge already at low temperatures ($T/S<0.1$) when compared to the half-quantized value. For example, $\lvert \kappa_{xy} \rvert/T = 9.414 \times 10^{-3} \, \mathrm{W}\mathrm{K}^2/\mathrm{m}$ or $\lvert \kappa_{xy}^\mathrm{2D} \rvert / T \approx 5 (\pi/6)(k_\mathrm{B}^2/\hbar)$ at $T/S=0.07$, roughly 10 times larger than the half-quantized value. As $T$ increases, $\lvert \kappa_{xy} \rvert/T$ grows and reaches a peak at $T/S=0.2$, beyond which it gradually decays, revealing a profile typical in the magnon thermal Hall conductivity. The maximum value of $\lvert \kappa_{xy} \rvert/T$ is about $20 \times 10^{-3} \, \mathrm{W} \mathrm{K}^2 / \mathrm{m}$, which is extraordinarily large, considering the appreciable signals in most thermal transport experiments is of the order $0.1 \times 10^{-3} \, \mathrm{W} \mathrm{K}^2 / \mathrm{m}$. Two factors work together to give a huge thermal Hall conductivity. (i) There are many low lying bands close to the zero energy. In particular, the excitation gap, i.e.~the gap between the lowest magnon band and the zero energy, is small. In these bands, the magnon occupation number is significant at low temperatures, resulting in sizable values of the $c_2$ function. (ii) The lowest $36$ bands in the magnon spectrum have the same Chern number $C_n=1$. The energy scale of the $36$th band is roughly $0.9 S$. For $T/S < 0.9$, the bands with index $n>36$ are not of much relevance as they are not thermally populated, resulting in small values of the $c_2$ function. On the other hand, by the approximation in \eqref{thermalhallconductivityformula}, the values of the $c_2$ function for the bands with index $n<36$ add up constructively as they have the same Chern number. In contrast, if the Chern number alternates between $+1$ and $-1$ as the band index $n$ increases, the values of $c_2$ function weighted by $C_n$ will tend to cancel out, leading to a small thermal Hall conductivity. Finally, we would like to comment on the temperature range in Fig.~\ref{thermalhalldata}. Readers should keep in mind that as the temperature approaches the intrinsic energy scale $\sim 1$, the data may no longer be meaningful because thermal fluctuations are likely to destroy the vortex liquid and results in a paramagnetic state with higher entropy.

\begin{figure}
\includegraphics[width=\linewidth]{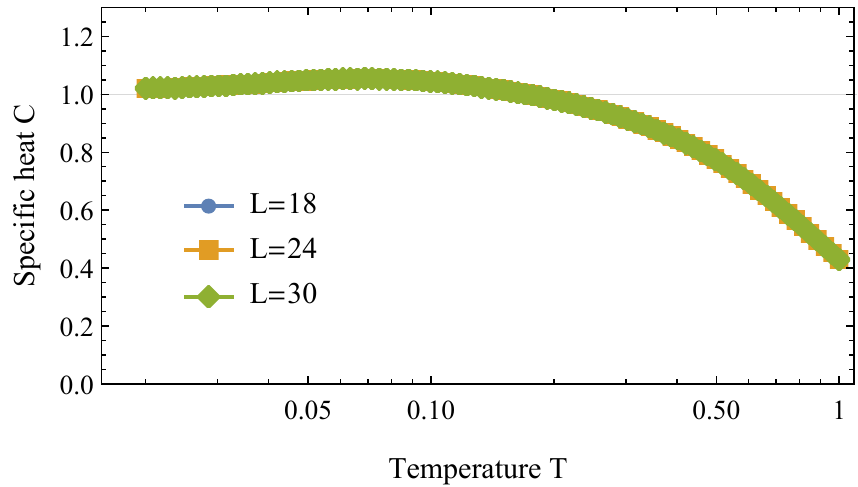}
\caption{The specific heat at $\varphi=\pi/4$ and $h/S=0.95$, for different system sizes $L=18, 24, 30$, as a function of temperature. The curves for all three system sizes coincide, indicating the absence of finite size effects in this regime. Statistical error bars are smaller than the plot markers.}
\label{fig:heat_B095}
\end{figure}

\subsection{\label{specificheat}Specific Heat}
With the magnetic vortex liquid being a hugely degenerate ensemble of vortex configurations at zero temperature, it resembles our understanding of a classical spin liquid. However, such an intricate degeneracy is often lifted by entropic corrections at finite temperatures, a mechanism known as \textit{order by disorder}\cite{Villain1980}. In order to investigate the stability of vortex liquid in the presence of thermal fluctuations, we perform finite temperature Monte Carlo simulations of the classical spin model \eqref{KGhspinmodel}. \\

We confirm that the vortex liquid exists as a thermal ensemble at finite temperatures. Also, in the parametric regime where a dilute vortex liquid is stabilized, the specific heat as a function of temperature does not show any sign of a thermal phase transition. The specific heat at $\varphi=\pi/4$ and $h/S=0.95$ is displayed in Fig.~\ref{fig:heat_B095} for three different system sizes. The data shows no dependence on the system size, which suggests the absence of finite size effect in the dilute vortex liquid. The transition into the vortex liquid state is rather a smooth crossover than a conventional phase transition. However, the finite extent of the system becomes relevant when the density of vortices increases. At $h/S \lesssim 0.9$ the specific heat acquires a system size dependence, while additional cusps indicates the onset of vortex crystallization (see Ref.~\onlinecite{SM}).

\subsection{\label{staticstructurefactor}Equal-Time Spin Correlator}
\begin{figure}
\includegraphics[width=\linewidth]{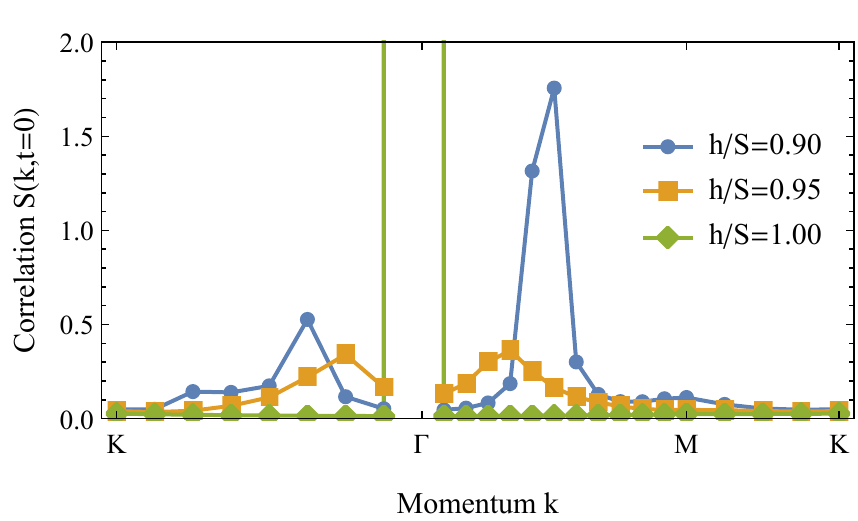}
\caption{The equal-time spin correlators at $\varphi=\pi/4$, $L=24$, and $T=0.02$, for different field strengths $h/S=0.9, 0.96, 1.0$. The intensity at the $\Gamma$ point extends up to $\mathcal{O}(10^3)$. A subleading intensity accumulates at a characteristic distance from the $\Gamma$ point which reflects the inter-vortex distance in real space.}
\label{fig:sf_equaltime}
\end{figure}

We would like to identify signatures of the magnetic vortex liquid from the two-spin correlation functions. First, we calculate the equal-time spin correlator
\begin{equation}
\mathcal{S}(\mathbf{k},t=0) = \frac{1}{N} \sum_{ij} e^{i \mathbf{k} \cdot (\mathbf{r}_j - \mathbf{r}_i)} \langle \mathbf{S}_i(0) \cdot \mathbf{S}_j(0) \rangle .
\end{equation}
We anticipate that at the Brillouin zone center (i.e.~the $\Gamma$ point), the dominant contribution scales with the total number of sites, since the majority of spins are polarized. Furthermore, we expect subleading peaks to emerge at some characteristic momentum scale $\lvert \mathbf{q}_\mathrm{vortex} \rvert$ which corresponds to the typical inter-vortex distance in real space. $\lvert \mathbf{q}_\mathrm{vortex} \rvert$ is expected to grow larger as the density of vortices increases and the inter-vortex spacing decreases. Indeed, as depicted in Fig.~\ref{fig:sf_equaltime}, such subleading peaks can be observed at $\varphi=\pi/4$ and $h/S=0.95$, whereas they disappear in the fully polarized state at $h/S=1.0$. These peaks can thus be conclusively attributed to the presence of magnetic vortices. \\

Furthermore, the enlargement of the characteristic scale $\lvert \mathbf{q}_\mathrm{vortex} \rvert$ is confirmed by comparison to a lower field $h/S=0.9$, where the vortices become dense. As we pointed out earlier, the vortices at higher densities are susceptible to forming a lattice structure. As a consequence of crystallization, the rotational symmetry in the equal-time spin correlator is broken, which is reflected in peaks with different heights along different paths in the Brillouin zone. In contrast, when the vortices are dilute (e.g.~at $h/S=0.95$), the rotational symmetry is preserved, resulting in peaks of equal height (see Fig.~\ref{fig:sf_equaltime}).

\begin{figure}
\includegraphics[width=\linewidth]{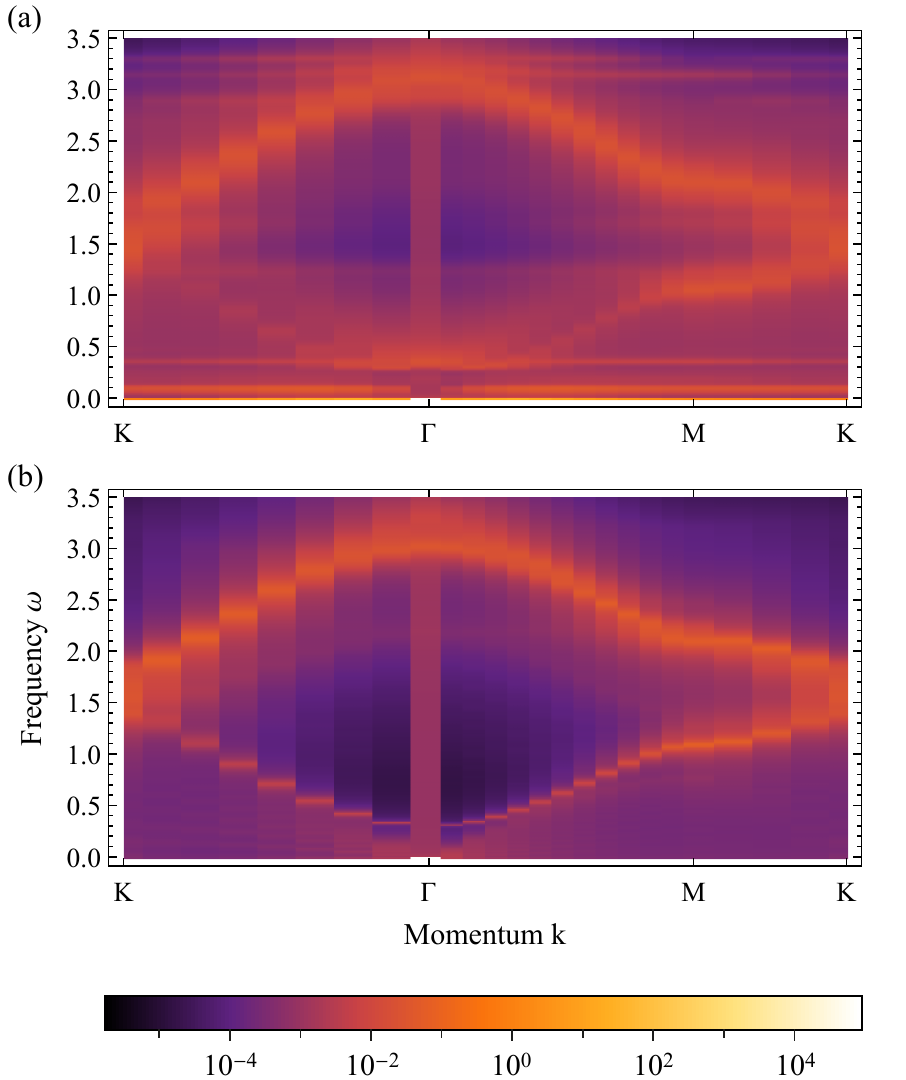}
\caption{The dynamic spin structure factors at $\varphi=\pi/4$, $L=24$, and $T=0.02$. (a) The field $h/S=0.95$ gives rise to a dilute vortex liquid. (b) In the fully polarized state at $h/S=1.0$ only the two conventional magnon bands are visible. Note that the leading $\omega=0$ divergence at the $\Gamma$ point gives rise to an unphysical tail at $\omega>0$, which is a numerical artifact of the Fourier transform over a finite time interval (see Sec.~\ref{finitetemperaturemontecarlo} for details).}
\label{fig:sf_dynamic}
\end{figure}

\subsection{\label{dynamicalstructurefactor}Dynamical Spin Structure Factor}
Finally, we compute the dynamic spin structure factor to extract information on the excitation spectrum. To achieve this, we consider spin configurations from the thermal ensemble generated in the Monte Carlo simulations and determine their time evolutions under the classical Landau-Lifshitz equations of motion. An animation of such a time evolution can be found in Ref.~\onlinecite{SM}. A subsequent Fourier transform leads to the dynamic spin structure factor
\begin{equation}
\mathcal{S}(\mathbf{k},\omega) = \frac{1}{2\pi N} \sum_{ij} \int \mathrm{d}t \, e^{i \omega t} e^{i \mathbf{k} \cdot (\mathbf{r}_j - \mathbf{r}_i)} \langle \mathbf{S}_i(t) \cdot \mathbf{S}_j(0) \rangle .
\end{equation}
As in the equal-time spin correlator, we expect dominant ferromagnetic correlations and the corresponding magnon dispersion (with two branches) in the dynamical spin structure factor. The presence of magnetic vortices further imprints subleading flat bands, as shown in Fig.~\ref{fig:sf_dynamic}a for $\varphi=\pi/4$ and $h/S=0.95$. The intensities from these flat bands smoothly blend into the background, resembling a continuum of spin excitations. For comparison, the dynamical structure factor of the fully polarized state is shown in Fig.~\ref{fig:sf_dynamic}b, which reveals two distinctive branches of magnon dispersion. \\

The dynamical spin structure factor calculated via the mean-field theory can be found in Ref.~\onlinecite{SM}. We find striking similarities between both calculations: (i) the emergence of a continuum of flat bands, (ii) the absence of signals at intermediate energies around the $\Gamma$ point, and (iii) stronger (weaker) signals around the upper (lower) branch of the magnon dispersion. This is because the vortex configuration and hence the flux distribution are random (under the constraint of the characteristic inter-vortex spacing), which yields an effectively uniform flux background when averaged across the thermal ensemble.

\section{\label{methods}Methods}
\subsection{\label{meanfieldmethod}Details of Mean-Field Theory}
Here we provide technical justifications to the mean-field approach in Sec.~\ref{meanfieldtheory}, which ultimately leads to an effective description of the vortex liquid as the bosonic analogue of Hofstadter problem. Since it involves multiple steps with different concepts, we organize the content in five subsections, as follows.

\subsubsection{\label{linearspinwavetheory}Linear Spin Wave Theory}
Our starting point is the application of linear spin wave theory\cite{PhysRev.58.1098,Jones_1987} to the field polarized state of \eqref{KGhspinmodel}, in which all spins are aligned in the $[111]$ direction. We perform a (global) coordinate transformation such that the $z$ axis of the new coordinate system is along the $[111]$ direction. Moreover, we fix the $x$ and $y$ directions to be $[11\bar{2}]$ and $[\bar{1}10]$ respectively. The spin $\tilde{\mathbf{S}}_i$ as measured in the new basis is related to $\mathbf{S}_i$ by $\mathbf{S}_i = R \tilde{\mathbf{S}}_i$, where
\begin{equation} \label{rotatetocrystal}
R = \begin{pmatrix} \frac{1}{\sqrt{6}} & - \frac{1}{\sqrt{2}} & \frac{1}{\sqrt{3}} \\ \frac{1}{\sqrt{6}} & \frac{1}{\sqrt{2}} & \frac{1}{\sqrt{3}} \\ -\sqrt{\frac{2}{3}} & 0 & \frac{1}{\sqrt{3}} \end{pmatrix} \in SO(3) .
\end{equation}
If we define
\begin{equation}
H_x = \begin{pmatrix} K & 0 & 0 \\ 0 & 0 & \Gamma \\ 0 & \Gamma & 0 \end{pmatrix}, \, H_y = \begin{pmatrix} 0 & 0 & \Gamma \\ 0 & K & 0 \\ \Gamma & 0 & 0 \end{pmatrix}, \, H_z = \begin{pmatrix} 0 & \Gamma & 0 \\ \Gamma & 0 & 0 \\ 0 & 0 & K \end{pmatrix},
\end{equation}
the spin Hamiltonian \eqref{KGhspinmodel} in the rotated basis is then $H = \sum_\lambda \sum_{\langle ij \rangle \in \lambda} \tilde{\mathbf{S}}_i^\mathrm{T} \tilde{H}_\lambda \tilde{\mathbf{S}}_j - \sum_i \tilde{\mathbf{h}} \cdot \tilde{\mathbf{S}}_i$, where $\tilde{H}_\lambda = R^\mathrm{T} H_\lambda R$ and $\tilde{\mathbf{h}} = R^\mathrm{T} \mathbf{h}$. We then carry out the Holstein Primakoff transformation,
\begin{subequations}
\begin{align}
\tilde{S}_i^z &= S - b_i^\dagger b_i = S - n_i , \label{holsteinprimakoffz} \\
\tilde{S}_i^x &= \frac{\sqrt{2S-n_i}b_i + b_i^\dagger \sqrt{2S-n_i}}{2} , \label{holsteinprimakoffx} \\
\tilde{S}_i^y &= \frac{\sqrt{2S-n_i}b_i - b_i^\dagger \sqrt{2S-n_i}}{2i} , \label{holsteinprimakoffy}
\end{align}
\end{subequations}
and keep only terms up to second order in $b$. At the rather special parametrization $K=-2\Gamma$, the rotated Hamiltonian components are, explicitly,
\begin{subequations}
\begin{align}
\tilde{H}_x &= \begin{pmatrix} - \Gamma & 0 & - \frac{\Gamma}{\sqrt{2}} \\ 0 & -\Gamma & \sqrt{\frac{3}{2}} \Gamma \\ -\frac{\Gamma}{\sqrt{2}} & \sqrt{\frac{3}{2}} \Gamma & 0 \end{pmatrix}, \label{rotatedxhamiltonian} \\
\tilde{H}_y &= \begin{pmatrix} - \Gamma & 0 & - \frac{\Gamma}{\sqrt{2}} \\ 0 & -\Gamma & - \sqrt{\frac{3}{2}} \Gamma \\ -\frac{\Gamma}{\sqrt{2}} & - \sqrt{\frac{3}{2}} \Gamma & 0 \end{pmatrix}, \label{rotatedyhamiltonian} \\
\tilde{H}_z &= \begin{pmatrix} - \Gamma & 0 & \sqrt{2} \Gamma \\ 0 & - \Gamma & 0 \\ \sqrt{2} \Gamma & 0 & 0 \end{pmatrix}, \label{rotatedzhamiltonian}
\end{align}
\end{subequations}
such that all magnon pairings vanish at the level of linear spin wave theory, yielding a Hamiltonian which is in essence the tight binding model of magnons \eqref{bosontightbinding}, where $\Gamma$ assumes the role of hopping integral and $h$ assumes the role of mass. Furthermore, \eqref{bosontightbinding} is exactly identical to the linear spin wave Hamiltonian of the XY model in a field along the $z$ axis, with the spin exchange $J=-\Gamma$. \\

We make a final remark concerning the nonvanishing matrix elements in \eqref{rotatedxhamiltonian}-\eqref{rotatedzhamiltonian} which connect the $x$ or $y$ component of one spin to the $z$ component of the other, e.g.~$-\Gamma/\sqrt{2}$ or $\sqrt{3/2} \Gamma$ in $\tilde{H}_x$. They will generate terms linear in $b$. However, as explained in Ref.~\onlinecite{Jones_1987}, the coefficients of these linear terms vanish, due to $\partial H / \partial \tilde{\mathbf{S}} \vert_{\tilde{\mathbf{S}}=(0,0,S)}$ in the classical ground state (or any local minimum).

\subsubsection{\label{continuummodel}Continuum Model}
We derive the continuum version of the Hamiltonian \eqref{KGhspinmodel} and, through a series of approximations, reduce it to the form of an XY model. For simplicity, we drop the Zeeman term as it does not affect the following analysis. \\

Assume that site $i$ belongs to sublattice $0$. Up to some constant, the Kitaev interaction can be rewritten as
\begin{equation} \label{kitaevcontinuum}
\begin{aligned}[b]
K \sum_{i \in 0} \sum_\lambda S_i^\lambda S_{i+\mathbf{n}_\lambda}^\lambda
&= - \frac{K}{2} \sum_\lambda \sum_{i \in 0} \left(S_i^\lambda - S_{i+\mathbf{n}_\lambda}^\lambda \right)^2 \\
&= - \frac{K}{2} \sum_\lambda \sum_{i \in 0} \Delta A \frac{\left(S_i^\lambda - S_{i+\mathbf{n}_\lambda}^\lambda \right)^2}{a^2} \\
&\longrightarrow -\frac{K}{2} \sum_\lambda \int \mathrm{d}A \, (\partial_{\mathbf{n}_\lambda}S_\mathbf{r}^\lambda)^2 ,
\end{aligned}
\end{equation}
where $\mathbf{n}_{\lambda=x,y,z}$ denote the three bond directions. This should only be taken as a shorthand notation though, because there are really just two independent directions in the two dimensional lattice. As will be argued later, the effective low energy theory is isotropic in the two dimensional space (resembling an XY model), so we should not worry too much about three $\mathbf{n}_\lambda$ at this point. One can show that the corresponding expression for $i \in 1$ is same as \eqref{kitaevcontinuum} due to the gradient term being squared. Similarly, the $\Gamma$ interaction can be rewritten as
\begin{equation} \label{gammacontinuum}
\begin{aligned}[b]
& \Gamma \sum_{i \in 0} \sum_\lambda \left( S_i^\mu S_{i+\mathbf{n}_\lambda}^\nu + S_i^\nu S_{i+\mathbf{n}_\lambda}^\mu \right) \\
& \longrightarrow - \Gamma \sum_\lambda \int \mathrm{d} A \, \left( \partial_{\mathbf{n}_\lambda} S_\mathbf{r}^\mu \partial_{\mathbf{n}_\lambda} S_\mathbf{r}^\nu - S_\mathbf{r}^\mu S_\mathbf{r}^\nu \right) ,
\end{aligned}
\end{equation}
where $(\lambda,\mu,\nu)$ is a cyclic permutation of $(x,y,z)$. Again, one can show that the corresponding expression for $i \in 1$ is same as \eqref{gammacontinuum}. We further neglect the on-site interaction $S_\mathbf{r}^\mu S_\mathbf{r}^\nu$ based on the following argument. In the classical limit $S \longrightarrow \infty$, we are allowed to symmetrize such a term as $(1/2) (S_\mathbf{r}^\mu S_\mathbf{r}^\nu + S_\mathbf{r}^\nu S_\mathbf{r}^\mu)$, but this becomes zero in the quantum limit $S = 1/2$ since different components of the spin operator anticommute. \\

Written more compactly, the continuum $K \Gamma$ model is thus
\begin{equation} \label{continuumKGmodel}
H = -\frac{1}{2} \sum_\lambda \int \mathrm{d} A \, \left( \partial_{\mathbf{n}_\lambda} \mathbf{S}_\mathbf{r} \right)^\mathrm{T} H_\lambda \left( \partial_{\mathbf{n}_\lambda} \mathbf{S}_\mathbf{r} \right) .
\end{equation}
Given a generic spin configuration $\lbrace \mathbf{S}_\mathbf{r} \rbrace$, let $R_\mathbf{r} \in SO(3)$ be the rotation matrix that defines the local coordinate frame, i.e.~$\mathbf{S}_\mathbf{r} = R_\mathbf{r} \tilde{\mathbf{S}}_\mathbf{r}$ and $\tilde{\mathbf{S}}_\mathbf{r} = (0,0,S)$. Let $\mathbf{S}_\mathbf{r}$ be parametrized by two angles $\theta_\mathbf{r}$ and $\phi_\mathbf{r}$ as $\mathbf{S}_\mathbf{r} = S (\sin \theta_\mathbf{r} \cos \phi_\mathbf{r}, \sin \theta_\mathbf{r} \sin \phi_\mathbf{r}, \cos \theta_\mathbf{r})$. We choose $R_\mathbf{r}$ to be\cite{Jones_1987,TATARA2019208}
\begin{equation} \label{localrotation}
R_\mathbf{r} = \begin{pmatrix} \cos \theta_\mathbf{r} \cos \phi_\mathbf{r} & - \sin \phi_\mathbf{r} & \sin \theta_\mathbf{r} \cos \phi_\mathbf{r} \\ \cos \theta_\mathbf{r} \sin \phi_\mathbf{r} & \cos \phi_\mathbf{r} & \sin \theta_\mathbf{r} \sin \phi_\mathbf{r} \\ - \sin \theta_\mathbf{r} & 0 & \cos \theta_\mathbf{r} \end{pmatrix} .
\end{equation}
We will make the spatial dependence of $\theta$ and $\phi$ implicit by dropping the subscript $\mathbf{r}$. It can be easily verified that \eqref{rotatetocrystal} corresponds to \eqref{localrotation} with $\theta = \cos^{-1} (1/\sqrt{3})$ and $\phi = \pi/4$. Switching to the local coordinate frame, \eqref{continuumKGmodel} becomes
\begin{equation} \label{localcontinuumKGmodel}
H = -\frac{1}{2} \sum_\lambda \int \mathrm{d} A \, \left[ \nabla_\lambda (\mathbf{r}) \tilde{\mathbf{S}}_\mathbf{r} \right]^\mathrm{T} \tilde{H}_\lambda (\mathbf{r}) \left[ \nabla_\lambda (\mathbf{r}) \tilde{\mathbf{S}}_\mathbf{r} \right] ,
\end{equation}
where $\nabla_\lambda (\mathbf{r}) = \partial_{\mathbf{n}_\lambda} + R_\mathbf{r}^\mathrm{T} \partial_{\mathbf{n}_\lambda} R_\mathbf{r}$ and $\tilde{H}_\lambda (\mathbf{r}) = R_\mathbf{r}^\mathrm{T} H_\lambda R_\mathbf{r}$. Since we are considering a textured ferromagnet, where most spins deviate only slightly from the $[111]$ direction, we replace the local Hamiltonian $\tilde{H}_\lambda (\mathbf{r})$ by a global one, $\tilde{H}_\lambda (\mathbf{r}) \approx R^\mathrm{T} H_\lambda R \equiv \tilde{H}_\lambda$ with $R$ given in \eqref{rotatetocrystal}. However, we still allow smooth spatial variation of $R_\mathbf{r}$ in $\nabla_\lambda (\mathbf{r})$. \\

At $K=-2\Gamma$, $\tilde{H}_\lambda$ is given by \eqref{rotatedxhamiltonian}-\eqref{rotatedzhamiltonian}. We further approximate the low energy physics of the textured ferromagnet as the XY model, i.e.~$\tilde{H}_\lambda \approx H_\mathrm{XY}=\mathrm{diag}(-\Gamma,-\Gamma,0)$. Heuristically, this is possible because the linear spin wave theory of the $K \Gamma$ model at such parametrization is equivalent to that of the XY model, which we justify as follows. For small deviations from the polarized state, following the discussion in the previous subsection, the contributions to the resulting linear spin wave Hamiltonian from the $S_i^x S_j^z$ and $S_i^y S_j^z$ terms are insignificant, which motivates us to set the corresponding matrix elements to $0$. This additional simplification allow us to treat the effect of inhomogeneous spin textures as a $U(1)$ gauge field, as discussed below.

\subsubsection{\label{gaugefield}Emergent Gauge Field}
We demonstrate that the inhomogeneous spin textures in the vortex liquid gives rise to a fictitious magnetic flux which will be experienced by the magnons. From \eqref{localrotation}, we have\cite{TATARA2019208}
\begin{equation} \label{localrotationderivative}
R^\mathrm{T} \partial_\mu R = \begin{pmatrix} 0 & 0 & 1 \\ 0 & 0 & 0 \\ -1 & 0 & 0 \end{pmatrix} \partial_\mu \theta + \begin{pmatrix} 0 & -\cos \theta & 0 \\ \cos \theta & 0 & \sin \theta \\ 0 & -\sin \theta & 0 \end{pmatrix} \partial_\mu \phi .
\end{equation}
Substituting \eqref{localrotationderivative} into \eqref{localcontinuumKGmodel} with $\tilde{H}_\lambda(\mathbf{r}) \approx H_\mathrm{XY}$, and performing the Holstein Primakoff transformation \eqref{holsteinprimakoffz}-\eqref{holsteinprimakoffy},
\begin{widetext}
\begin{equation}
\begin{aligned}[b]
\frac{H}{S} &= \Gamma \sum_\lambda \int \mathrm{d} A \, \left[ \partial_{\mathbf{n}_\lambda} b_\mathbf{r}^\dagger \partial_{\mathbf{n}_\lambda} b_\mathbf{r} - i \cos \theta \partial_{\mathbf{n}_\lambda} \phi \left(b_\mathbf{r}^\dagger \partial_{\mathbf{n}_\lambda} b_\mathbf{r} - b_\mathbf{r} \partial_{\mathbf{n}_\lambda} b_\mathbf{r}^\dagger \right) \right] + O \left[ \left( \partial_{\mathbf{n}_\lambda} R \right)^2 \right] \\
&\approx \Gamma \sum_\lambda \int \mathrm{d} A \, [ \left( \partial_{\mathbf{n}_\lambda} + i A_{\mathbf{n}_\lambda} \right) b_\mathbf{r}^\dagger ] [ \left( \partial_{\mathbf{n}_\lambda} - i A_{\mathbf{n}_\lambda} \right) b_\mathbf{r} ],
\end{aligned}
\end{equation}
\end{widetext}
where we have neglected terms quadratic in the derivative of the local spin structure, as well as terms linear in $b$ (which describe the interaction between magnons and the local spin structure)\cite{TATARA2019208}. In the second equality, the integrand has the same expression in all bond directions $\lambda=x,y,z$, so we may as well replace $\mathbf{n}_\lambda$ by $\mu=x,y$. We see that the effect of local rotation enters as a $U(1)$ gauge field $A_\mu = - \cos \theta \partial_\mu \phi$ that couples to the magnons, which is twice as large as that for conduction electrons hopping in a nonuniform spin background\cite{PhysRevB.91.104435,nnano.2013.243,1.4870695}. In other words, the magnons experience a fictitious magnetic flux due to the $U(1)$ gauge field which originates from the spatial variation of spins in the vortex liquid.

\subsubsection{\label{uniformflux}Uniform Flux Approximation}
\begin{figure}
\subfloat[]{\label{partitionone}
\includegraphics[scale=0.25]{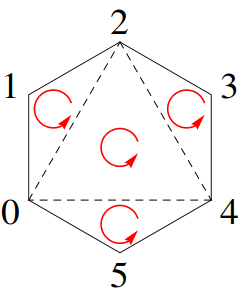}} \qquad \qquad
\subfloat[]{\label{partitiontwo}
\includegraphics[scale=0.25]{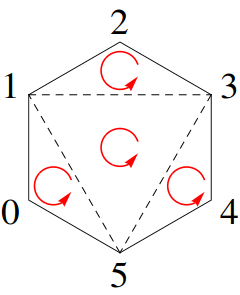}}
\caption{(a) and (b) are two examples of partitioning the unit hexagon into four triangles. The summation of solid angles subtended by the spins on the triangles under different partitionings is equal up to an additive factor of $4 \pi$, which has no physical consequence. The arrows define the orientation of the spin chirality in \eqref{solidangleformula}.}
\end{figure}

To calculate the fictitious magnetic flux $\phi$ of some area $A$, we take the curl of the gauge field obtained in the previous subsection
\begin{equation}
(\nabla \times \mathbf{A})_\lambda = \frac{1}{2} \varepsilon_{\lambda \mu \nu} \mathbf{S} \cdot \partial_\mu \mathbf{S} \times \partial_\nu \mathbf{S}
\end{equation}
and integrate it over $A$, which is equal to the solid angle subtended by the spins around the boundary $\partial A$\cite{PhysRevB.101.024420}. Returning to the discrete model (i.e.~lattice), we want to calculate the flux penetrating each unit hexagon. To achieve this, we partition each unit hexagon into four triangles (see Figs.~\ref{partitionone} and \ref{partitiontwo} for instance), and add up the solid angles of these triangles according to the formula\cite{BERG1981412,4121581}
\begin{equation} \label{solidangleformula}
\frac{\Omega_{ijk}}{2} = \mathrm{arg} \left(1 + \frac{\mathbf{S}_i \cdot \mathbf{S}_j + \mathbf{S}_j \cdot \mathbf{S}_k + \mathbf{S}_k \cdot \mathbf{S}_i}{S^2} + i \frac{\mathbf{S}_i \cdot \mathbf{S}_j \times \mathbf{S}_k}{S^3} \right) .
\end{equation}
Imagine that a magnon hops around these triangles, say, in the anticlockwise sense, the $U(1)$ phases gained along the internal lines will cancel (up to an additive factor of $2 \pi$) as they are travelled exactly once along opposite directions. The summation of solid angles under different partitionings of the unit hexagon differ by an integer multiple of $4 \pi$, which has no physical consequence ($e^{4 \pi i}=1$). \\

We fix the range of flux per hexagon to be $[-2 \pi, 2 \pi)$. One may argue that any interval of length $4 \pi$, for example $[0, 4\pi)$, are equally valid, which is true if we do not average the flux over the lattice. Since we are planning elsewise, we justify why the choice of $[-2 \pi, 2 \pi)$ is more physical than others as follows. \\

The total flux (i.e.~summation of the fluxes of all hexagons) is an integer multiple of $2 \pi$, which is consistent with the $U(1)$ gauge theory. This is because each nearest bond is travelled exactly once along opposite directions. By restricting the flux per hexagon between $-2 \pi$ and $2 \pi$ when we are evaluating \eqref{solidangleformula}, the total flux is equal to $- 4 \pi$ multiplied by the total number of vortices, a feature which does not generally hold for other intervals. Also, the spatial gradient of the local spin structure is small in a textured ferromagnet, so the fluxes of most unit hexagons should be close to zero. It makes more sense to assign to a unit hexagon a flux of $+\epsilon$ or $-\epsilon$, rather than $\epsilon$ or $4 \pi - \epsilon$ for instance, where $\epsilon$ is some small positive real number. We emphasize again that we have to carefully choose the range of flux per unit hexagon only because we are going to average the flux over the lattice. If we were not to do so, then we would be satisfied with any interval of length $4 \pi$. \\

Suppose that there are $n \in \mathbb{N}$ vortices such that the total flux of the vortex liquid is $- 4 \pi n$, and the system has a total number of $N$ unit hexagons. We then spread the total flux uniformly over the system, so the flux of each unit hexagon is $- 4 \pi n / N = - 2 \pi p / q$, where $p$ and $q$ are relatively prime. This is reminiscent of the famous Hofstadter problem in which electrons hopping on a lattice are subjected to a uniform magnetic field, as discussed in the Sec.~\ref{meanfieldtheory}.

\subsubsection{\label{magnonspectrummethod}Magnon Energy Spectrum}
After averaging the fictitious magnetic flux, we attach the $U(1)$ gauge field (of the optimal gauge) to the magnons in the polarized state, which yields the bosonic Hofstadter model. Since a series of approximations have been made, we have to slightly tune the magnetic field $h$ in \eqref{hofstadtermagnon} or we risk getting an unstable spectrum with negative energies. For instance, we cannot use $h/S=1.3$ directly in \eqref{hofstadtermagnon} for $\phi=-2 \pi \times 1 / 50$, even though the corresponding spin configuration is obtained at $h/S=1.3$. To resolve this issue, we set $h$ to be the critical field $h_\mathrm{crit}=3 \lvert \Gamma \rvert$ at which the system is fully polarized, which is a reasonable choice because the vortex liquid takes place near the polarized regime. At the critical field, the zero flux magnon spectrum (i.e.~the simple polarized state neglecting any effect of the inhomogeneous spin texture) is exactly gapless, but a finite flux like $\phi=-2 \pi \times 1/50$ induces a small gap stabilizing the magnon spectrum.

%Applying the Fourier transform $b_{i,n} = (1/\sqrt{N}) b_{\mathbf{k},n} e^{i \mathbf{k} \cdot \mathbf{R}_i}$, where $i$ labels the magnetic unit cell and $n=1, \ldots, 2 q$ labels the sublattice within the magnetic unit cell, the linear spin wave Hamiltonian \eqref{bosontightbinding} can be cast into the form $H = \sum_\mathbf{k} \Psi_\mathbf{k}^\dagger D_\mathbf{k} \Psi_\mathbf{k}$, where $\Psi_\mathbf{k} = (b_{\mathbf{k},1}, \ldots, b_{\mathbf{k},2q})$ and $D_\mathbf{k}$ is a $2q$ dimensional Hermitian matrix. Since there is no magnon pairing, a unitary transformation $U_\mathbf{k}$ is sufficient to diagonalize $D_\mathbf{k}$ while preserving the bosonic commutation relation between magnons (a Bogoliubov transformation is required if magnon pairing is present). We finally arrives $H=\sum_{n\mathbf{k}} \varepsilon_{n\mathbf{k}} \gamma^\dagger_{\mathbf{k},n} \gamma_{\mathbf{k},n}$, where $\gamma$ is defined through $\Psi_\mathbf{k} = U_\mathbf{k} \Gamma_\mathbf{k}$ with $\Gamma_\mathbf{k}=(\gamma_{\mathbf{k},1}, \ldots, \gamma_{\mathbf{k},2q})$. $\lbrace \epsilon_{n\mathbf{k}} \rbrace$ is the magnon spectrum.

\subsection{\label{chernnumber}Evaluation of Chern Number}
As shown in \eqref{thermalhallconductivityformula}, we need the Chern numbers of the magnon bands to calculate the thermal Hall conductivity. Here we elaborate on how to evaluate the Chern number for a given band. \\

If the $n$th band is nondegenerate, the Berry connection, Berry curvature, and Chern number are given respectively as\cite{rspa.1984.0023,1.4870695,JPSJ.74.1674}
\begin{subequations}
\begin{align}
A_{\mu, n \mathbf{k}} &= i \langle n \mathbf{k} \lvert \partial_{k_\mu} \rvert n \mathbf{k} \rangle , \label{berryconnectionformula} \\
\Omega_{n \mathbf{k}} &= \partial_{k_\mu} A_{\nu, n \mathbf{k}} - \partial_{k_\nu} A_{\mu, n \mathbf{k}} , \label{berrycurvatureformula} \\
C_n &= \frac{1}{2\pi} \int_\mathrm{FBZ} \mathrm{d}^2 \mathbf{k} \, \Omega_{n \mathbf{k}} \longrightarrow \frac{1}{2\pi} \sum_{\mathbf{k} \in \mathrm{FBZ}} \frac{(2\pi)^2}{A} \Omega_{n \mathbf{k}}, \label{chernnumberformula}
\end{align}
\end{subequations}
where $\lvert n\mathbf{k} \rangle$ is the (normalized) eigenstate corresponding to the eigenvalue $\varepsilon_{n \mathbf{k}}$, $\mu,\nu=x,y$, FBZ denotes the first Brillouin zone, and in the last equality we have switched from the continuum integral to the discrete summation with $A$ denoting the total area of the system (not to be confused with the Berry connection or the gauge field). \\

If the $n$th band is $M$-fold degenerate, the Chern number is defined collectively for the degenerate manifold as follows. We first define the multiplet $\psi_{\mathbf{k}} = ( \lvert n_1 \mathbf{k} \rangle, \ldots, \lvert n_M \mathbf{k} \rangle)$. The (nonabelian) Berry connection, Berry curvature and Chern number are given respectively as\cite{JPSJ.73.2604,JPSJ.74.1374,JPSJ.74.1674}
\begin{subequations}
\begin{align}
A_\mathbf{k} &= i \psi_\mathbf{k}^\dagger \mathrm{d} \psi_\mathbf{k}, \label{nonabelianberryconnectionformula} \\
\Omega_\mathbf{k} &= \mathrm{d} A_\mathbf{k}, \label{nonabelianberrycurvatureformula} \\
C_\psi &= \frac{1}{2\pi} \int_\mathrm{FBZ} \mathrm{Tr} \, \Omega_\mathbf{k} \longrightarrow \frac{1}{2 \pi} \sum_{\mathbf{k} \in \mathrm{FBZ}} \frac{(2 \pi)^2}{A} \mathrm{Tr} \, \Omega_\mathbf{k}, \label{nonabelianchernnumberformula}
\end{align}
\end{subequations}
where $\mathrm{d}$ is the differential operator such that $A_\mathbf{k}$ is a differential 1-form and $\Omega_\mathbf{k}$ is a differential 2-form. \\

For $\phi=-2 \pi \times 1/50$ describing the vortex liquid at $\varphi=\tan^{-1} (1/2)$ and $h/S=1.3$, we check the convergence of each Chern number to an integer with increasing momentum grids in the first Brillouin zone, and that the summation of all Chern numbers is zero, $\sum_{n=1}^{2q} C_n = 0$. We find that $72$ out of $2q=100$ bands in the magnon spectrum (see Fig.~\ref{dispersion}) carry Chern number $1$, while the largest Chern number is $51$. None of the bands have Chern number $0$, signifying their topological nontriviality. \\

While most of the bands at low and high energies are nondegenerate, some bands at intermediate energies (roughly between $\omega/S=1$ and $2$) are two fold degenerate. Strictly speaking, the formula \eqref{thermalhallconductivityformula} is derived only for nondegenerate bands, but the second equality in \eqref{thermalhallconductivityformula} suggests that we can treat an $M$-fold degenerate band as a single band with a collective Chern number $C_\psi$. \\

Finally, we remark on the how the Chern numbers change when we invert the magnetic field. When the field is along the $[111]$ direction, the ground state spin configuration is $\lbrace \mathbf{S}_i \rbrace$, the flux carried by each vortex is negative, so is the total flux. The corresponding set of Chern numbers $\lbrace C_n \rbrace$ yields a negative thermal Hall conductivity. When the field is along the $[\bar{1}\bar{1}\bar{1}]$ direction, the ground state spin configuration changes as $\lbrace \mathbf{S}_i \rbrace \longrightarrow \lbrace -\mathbf{S}_i \rbrace$. Meanwhile, the linear spin wave Hamiltonian of the background polarized state under the $[\bar{1}\bar{1}\bar{1}]$ field is the same as that under the $[111]$ field. Therefore, we still obtain the vortex liquid under $\mathbf{h} \longrightarrow -\mathbf{h}$ as long as the field magnitude remains the same, yet each vortex has an opposite chirality - the flux carried by each vortex is now positive, and the resulting set of Chern numbers changes as $\lbrace C_n \rbrace \longrightarrow \lbrace - C_n \rbrace$. By \eqref{thermalhallconductivityformula}, the thermal Hall conductivity flips sign and is now positive.

\subsection{\label{finitetemperaturemontecarlo}Finite Temperature Monte Carlo Simulation}
In order to access the thermal properties of the classical spin model defined in \eqref{KGhspinmodel}, we resort to finite temperature Monte Carlo simulations. We perform the simulations on lattices with $L\times L$ unit cells (or $N=L \times L \times 2$ sites) with periodic boundary conditions, typically at $L=24$. We equilibrate the simulations for $10^7$ sweeps, where a single sweep amounts to $N$ attempted local spin updates, before taking measurements for an additional $10^8$ sweeps. To improve the convergence of the simulation, we employ a parallel tempering technique to simultaneously equilibrate systems at $96$ different temperatures, which are logarithmically spaced between $T_\mathrm{min}=0.02$ and $T_\mathrm{max}=1$. \\

For the calculation of the spin correlations and the dynamic structure factor, we process one spin configuration every $2,000$ sweeps, i.e. a total of $50,000$ configurations generated throughout the Monte Carlo simulation. The time evolution of every individual spin configuration is obtained by solving the Landau-Lifshitz equation\cite{Landau1935,Lakshmanan2011} 
\begin{equation} \label{landaulifshitz}
\begin{aligned}[b]
\frac{\mathrm{d}S_i^x}{\mathrm{d}t} = & K \left( S_i^z S_{i_y}^y - S_i^y S_{i_z}^z \right) + \Gamma \left( S_i^z S_{i_x}^z + S_i^z S_{i_z}^z - S_i^y S_{i_x}^y - S_i^y S_{i_y}^x \right) \\
& +\frac{h}{\sqrt{3}} \left( S_i^y - S_i^z \right)
\end{aligned}
\end{equation}
(with cyclic permutations of $x$, $y$, and $z$), where $i_\lambda$ indicates the nearest neighbor connected to $i$ via the $\lambda \in \lbrace x,y,z \rbrace$ bond. \eqref{landaulifshitz} is solved numerically up to a time $t_\mathrm{max}=500$, with a dynamic step size and a local error tolerance chosen to produce convergent results at least up to $t_\mathrm{max}$\cite{rackauckas2017differentialequations,rackauckas2019confederated}. The structure factor is then obtained by Fourier transforming the time evolution of spins (in steps of $\Delta t=0.05$) and subsequent thermal averaging over all $50,000$ configurations.

\section{\label{materialrelevance}Relevance to Real Materials}
There exists a self-dual transformation $\mathcal{T}_1$ of the $J K \Gamma \Gamma'$ model, under which all spins are rotated uniformly by $\pi$ about the $[111]$ direction\cite{PhysRevB.92.024413}. Here we apply $\mathcal{T}_1$ to the $K \Gamma$ model with $K>0$ and $\Gamma<0$ investigated in this work. The parameters in the rotated frame are given by $\tilde{J} =  4 K / 9 - 4 \Gamma / 9 , \tilde{K} = - K / 3 + 4 \Gamma / 3 , \tilde{\Gamma} = 4 K / 9 + 5 \Gamma / 9$, and $\tilde{\Gamma}' = - 2 K / 9 + 2 \Gamma / 9$, such that the sign of the Kitaev interaction $\tilde{K}$ is always negative, while that of $\tilde{\Gamma}$ is positive for $\lvert \Gamma / K \rvert < 4/5$. Such a sign structure is the same as that in $\alpha$-RuCl$_3$\cite{PhysRevB.93.155143,PhysRevB.93.214431,PhysRevB.96.115103,s41467-017-01177-0,s41567-020-0874-0} and similar materials. Meanwhile, the $[111]$ field remains invariant under $\mathcal{T}_1$ since it lies exactly along the rotation axis. \\

\begin{figure}
\includegraphics[scale=0.25]{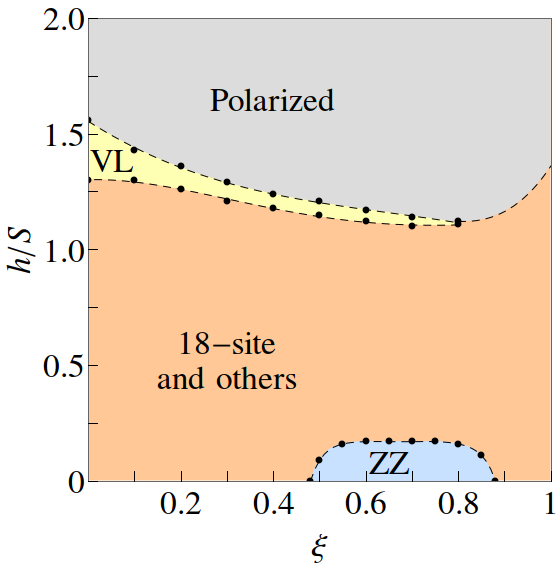}
\caption{\label{extent}The dual model of the $K \Gamma$ model with $K>0$ and $\Gamma<0$ is continuously deformed to models that are relevant to $\alpha$-RuCl$_3$ and similar materials, by tuning $\xi$ from $0$ to $1$. There is a significant extent of the magnetic vortex liquid to large values of $\xi$, where the ground state in the low field regime is the ZZ order.}
\end{figure}

$\alpha$-RuCl$_3$ has dominant $\tilde{K}<0$ and $\tilde{\Gamma}>0$ interactions with $\lvert \tilde{K} / \tilde{\Gamma} \rvert \approx 2$\cite{PhysRevB.93.155143,PhysRevB.96.115103,s41467-017-01177-0}. On top of these, other interactions such as $\tilde{J}$, $\tilde{\Gamma}'$ and/or $\tilde{J}_3$ at the subleading order are present, so that the zigzag (ZZ) magnetically ordered state is stabilized in the zero field limit, as observed in experiments\cite{PhysRevB.91.144420,PhysRevB.92.235119}. We investigate the extent of the magnetic vortex liquid towards more realistic model parameters, as follows. First, we choose $\lvert K/\Gamma \rvert = 22/5$ (or $\varphi = \tan^{-1} (5/22) \approx 0.07 \pi$) in \eqref{KGhspinmodel} such that $\lvert \tilde{K}/\tilde{\Gamma} \rvert = 2$ in the dual model. However, the corresponding $\tilde{J}$ and $\tilde{\Gamma}'$ are quite significant. We then continuously deform the dual model to one in which $\tilde{J}$ and $\tilde{\Gamma}'$ vanish. More precisely, we introduce a tunable parameter $\xi \in [0,1]$ such that (i) at $\xi=0$, the set of parameters $\lbrace \tilde{J},\tilde{K},\tilde{\Gamma},\tilde{\Gamma}' \rbrace$  is mapped from \eqref{KGhspinmodel} with $\varphi=\tan^{-1} (5/22)$ under $\mathcal{T}_1$, (ii) at $\xi=1$, $\tilde{J}=0$, $\tilde{\Gamma}'$=0, and $\sqrt{\tilde{K}^2+\tilde{\Gamma}^2}=1=\sqrt{K^2+\Gamma^2}$, and (iii) the ratio $\lvert \tilde{K} / \tilde{\Gamma} \rvert=2$ is kept fixed throughout the deformation. The Hamiltonian describing the deformation can be found in Appendix~\ref{materialrelevanceappendix}. We indicate the extent of vortex liquid as $\xi$ is increased from $0$ to $1$ in Fig.~\ref{extent}. \\

Remarkably, the vortex liquid survives up to $\xi \approx 0.8$. Although the structure of vortex greatly simplifies beyond $\xi \approx 0.2$, the fluid-like nature remains - the vortices do not form a crystalline order, but they appear at rather random positions in the background of polarized spins\cite{SM}. On the other hand, the ground state at zero and low fields for $0.5 \lesssim \xi \lesssim 0.9$ is the ZZ order. Considering the set of parameters for $\xi \gtrsim 0.5$ as realistic, our result suggests that the vortex liquid may be relevant to real materials like $\alpha$-RuCl$_3$ at high fields.

\section{\label{discussion}Discussion}
Considering frustrated honeycomb magnets with bond-dependent interactions, we unveil the magnetic vortex liquid state in the presence of an external magnetic field. The magnetic vortices are textures made out of a large number of spins and embedded in the background of polarized spins. Instead of settling in a crystalline order, the vortices form a liquid state, while distinct configurations of the vortices form a thermal ensemble. We show that this novel phase leads to a continuum of spin excitations that can be seen in neutron scattering experiments and a large thermal Hall conductivity. \\

Furthermore, we investigate the relevance of the vortex liquid to existing Kitaev materials like $\alpha$-RuCl$_3$ by continuously evolving the dual model of \eqref{KGhspinmodel} towards more realistic models\cite{PhysRevB.93.155143,PhysRevB.96.115103,s41467-017-01177-0}. We establish the survival of the vortex liquid at realistic interaction parameters, where the ground state in the low field regime is the ZZ order. Our result suggests the possible existence of the vortex liquid phase in real materials. However, the highest magnetic fields (e.g.~$\sim$ 60 T in Ref.~\onlinecite{PhysRevB.92.235119}) that can be applied in experiments have so far failed to polarize $\alpha$-RuCl$_3$ in the $[111]$ direction. One may look for other Kitaev materials with spin interactions at a lower energy scale, such as $f$-electron based honeycomb magnets\cite{PhysRevB.99.241106}, so that the polarized state and hence the vortex liquid are practically accessible under a $[111]$ field. \\

Nonetheless, it is interesting to see that the spin excitation continuum and the large thermal Hall conductivity discovered in the magnetic vortex liquid state are reminiscent of the experimental observations in $\alpha$-RuCl$_3$. There is an important difference, though, between our work and established experiments, in terms of field direction. Most of the relevant experiments on $\alpha$-RuCl$_3$ were carried out under in-plane and tilted magnetic fields. That is, the neutron scattering continuum is observed in the presence of a magnetic field parallel to the honeycomb plane\cite{PhysRevLett.120.077203,s41535-018-0079-2,PhysRevB.100.060405}, while the half-quantized thermal Hall conductivity was discovered in magnetic fields tilted away from the $[111]$ direction and along certain in-plane direction\cite{s41586-018-0274-0,2001.01899}. Our model, on the other hand, would be more relevant to future experiments with the $[111]$ magnetic field. \\

Finally, we discuss the relation of our work to existing theoretical studies. A classical phase diagram in the low field regime $h/S \in [0, 0.2]$ of the same model \eqref{KGhspinmodel} has recently been reported in Ref.~\onlinecite{2004.14415}, which used an unsupervised machine learning method. The magnetic vortex liquid is not relevant at such low fields as it only occupies a tiny area. Moreover, it may be difficult for unsupervised machine learning to give an appropriate interpretation to an inhomogeneous state in which the majority of spins are still polarized. These are the possible reasons that the vortex liquid was not identified in Ref.~\onlinecite{2004.14415}. \\

On the other hand, the quantum model of \eqref{KGhspinmodel} has recently been investigated using the density matrix renormalization group (DMRG) on the two-leg ladder system\cite{2007.07259}. Among the plethora of phases presented in the phase diagram, the uniform chirality (UC) phase and the staggered chirality (SC) phase, where the scalar spin chirality is finite, could be related to the vortex liquid in our work. This is because the noncoplanar spin structure of each vortex gives rise to a finite scalar spin chirality. The precise connection, however, is not clear at the moment. The fate of the vortex liquid in the quantum model will be an important and interesting subject of future study.

\begin{acknowledgments}
We thank Moon Jip Park and Hae-Young Kee for useful discussions. L.E.C.~was supported by the Ontario Graduate Scholarship. Y.B.K.~was supported by the Killam Research Fellowship from the Canada Council for the Arts, the NSERC of Canada, and the Center for Quantum Materials at the University of Toronto. Most of the computations were performed on the Cedar and Niagara clusters, which are hosted by WestGrid and SciNet\cite{1742-6596-256-1-012026} in partnership with Compute Canada. \\

%L.E.C.~did the analysis at zero temperature and calculated the thermal Hall conductivity. F.L.B.~did the analysis at finite temperatures and calculated the static and dynamical spin structure factors. Y.B.K.~conceived and supervised the study. All authors contributed to the writing of the manuscript.
\end{acknowledgments}

\appendix

\section{\label{materialrelevanceappendix}Towards Realistic Models}
In Sec.~\ref{materialrelevance}, we investigate the extent of magnetic vortex liquid towards the parametric regime relevant to $\alpha$-RuCl$_3$ by introducing a continuous deformation parametrized by $\xi \in [0,1]$,
\begin{widetext}
\begin{equation}\label{homeomorphism}
H (\xi) = (1-\xi) H \left( \tilde{J}=\frac{12}{\sqrt{509}},\tilde{K}=\frac{-14}{\sqrt{509}},\tilde{\Gamma}=\frac{7}{\sqrt{509}},\tilde{\Gamma}'=\frac{-6}{\sqrt{509}} \right) + \xi H \left( \tilde{J}=0,\tilde{K}=\frac{-2}{\sqrt{5}},\tilde{\Gamma}=\frac{1}{\sqrt{5}},\tilde{\Gamma}'=0 \right) - \sum_i \mathbf{h} \cdot \mathbf{S}_i .
\end{equation}
\end{widetext}
$H(\xi=0)$ is dual to the $K \Gamma$ model with $K=22/\sqrt{509}$ and $\Gamma=-5/\sqrt{509}$, while $H(\xi=1)$ is the $\tilde{K} \tilde{\Gamma}$ model with $\tilde{K}=-2/\sqrt{5}$ and $\tilde{\Gamma}=1/\sqrt{5}$. The Zeeman term with $\mathbf{h}$ along the $[111]$ direction does not depend on $\xi$. Explicitly, the interaction parameters $\tilde{J} (\xi)$, $\tilde{K} (\xi)$, $\tilde{\Gamma} (\xi)$, and $\tilde{\Gamma}' (\xi)$ in \eqref{homeomorphism} are given by
\begin{subequations}
\begin{align}
\tilde{J} (\xi) &= (1-\xi) \frac{12}{\sqrt{509}}, \\
\tilde{K} (\xi) &= -(1-\xi) \frac{14}{\sqrt{509}} -\xi \frac{2}{\sqrt{5}}, \\
\tilde{\Gamma} (\xi) &= (1-\xi) \frac{7}{\sqrt{509}} +\xi \frac{1}{\sqrt{5}}, \\
\tilde{\Gamma}' (\xi) &= -(1-\xi) \frac{6}{\sqrt{509}} .
\end{align}
\end{subequations}
We display the values of these parameters, normalized by $\lvert \tilde{K} (\xi) \rvert$, at various $\xi$ in Table.~\ref{xiparameter}. We also show examples of spin configurations of the vortex liquids at various $\xi$ in Ref.~\onlinecite{SM} (see Figs.~S4a, S4b, S5a, and S5b).

\begin{table}
\caption{\label{xiparameter}The interaction parameters of $H (\xi)$ at various $\xi$.}
\begin{tabular}{|>{\centering\arraybackslash}m{20 pt}|>{\centering\arraybackslash}m{50 pt}|>{\centering\arraybackslash}m{50 pt}|>{\centering\arraybackslash}m{50 pt}|>{\centering\arraybackslash}m{50 pt}|}
\hline
$\xi$ & $\tilde{J} (\xi) / \lvert \tilde{K} (\xi) \rvert$ & $\tilde{K} (\xi) / \lvert \tilde{K} (\xi) \rvert$ & $\tilde{\Gamma} (\xi) / \lvert \tilde{K} (\xi) \rvert$ & $\tilde{\Gamma}' (\xi) / \lvert \tilde{K} (\xi) \rvert$ \\ \hline
$0$	& $0.857$ & $-1$ & $0.5$ & $-0.429$ \\
$0.1	$ & $0.739$ & $-1$ & $0.5$ & $-0.369$ \\ 
$0.2	$ & $0.630$ & $-1$ & $0.5$ & $-0.315$ \\
$0.3	$ & $0.530$ & $-1$ & $0.5$ & $-0.265$ \\
$0.4	$ & $0.437$ & $-1$ & $0.5$ & $-0.219$ \\
$0.5	$ & $0.351$ & $-1$ & $0.5$ & $-0.176$ \\
$0.6	$ & $0.271$ & $-1$ & $0.5$ & $-0.136$ \\
$0.7	$ & $0.196$ & $-1$ & $0.5$ & $-0.098$ \\
$0.8	$ & $0.127$ & $-1$ & $0.5$ & $-0.063$ \\
$0.9$ & $0.061$ & $-1$ & $0.5$ & $-0.031$ \\
$1$	& $0$ & $-1$ & $0.5$ & $0$ \\
\hline
\end{tabular}
\end{table}

\bibliography{reference200821}

\clearpage

\onecolumngrid

\begin{center}
\textbf{\large Supplementary Materials: \\ Magnetic vortex liquid and anomalous thermal Hall conductivity \\ in frustrated magnets with bond-dependent interactions}
\end{center}
\begin{center}
Li Ern Chern$^1$, Finn Lasse Buessen$^1$, and Yong Baek Kim$^1$
\end{center}
\begin{center}
{\small
\textit{$^1$Department of Physics, University of Toronto, Toronto, Ontario M5S 1A7, Canada}}
\end{center}

\setcounter{equation}{0}
\setcounter{figure}{0}
\setcounter{table}{0}
\setcounter{page}{1}

\renewcommand{\thesection}{S\arabic{section}}
\renewcommand{\theequation}{S\arabic{equation}}
\renewcommand{\thefigure}{S\arabic{figure}}
\renewcommand{\thetable}{S\arabic{table}}

\section*{S1.~Real Space Spin Configurations}
We show more examples of spin configurations of the magnetic vortex liquid states at various parameters $\varphi$ and $h$, in Figs.~S1a, S1b, S2a, and S2b. Fig.~S3 is the vortex liquid to which we apply the mean field theory in Sec.~III C. \\

In Sec.~V and Appendix A, we introduce a parameter $\xi \in [0,1]$ that continuously deforms the original $K \Gamma$ model towards more realistic models. We show examples of spin configurations of the vortex liquids at various $\xi$ in Figs.~S4a, S4b, S5a, and S5b.

\begin{figure}
\subfloat[]{\label{phi025h092}
\includegraphics[scale=0.36]{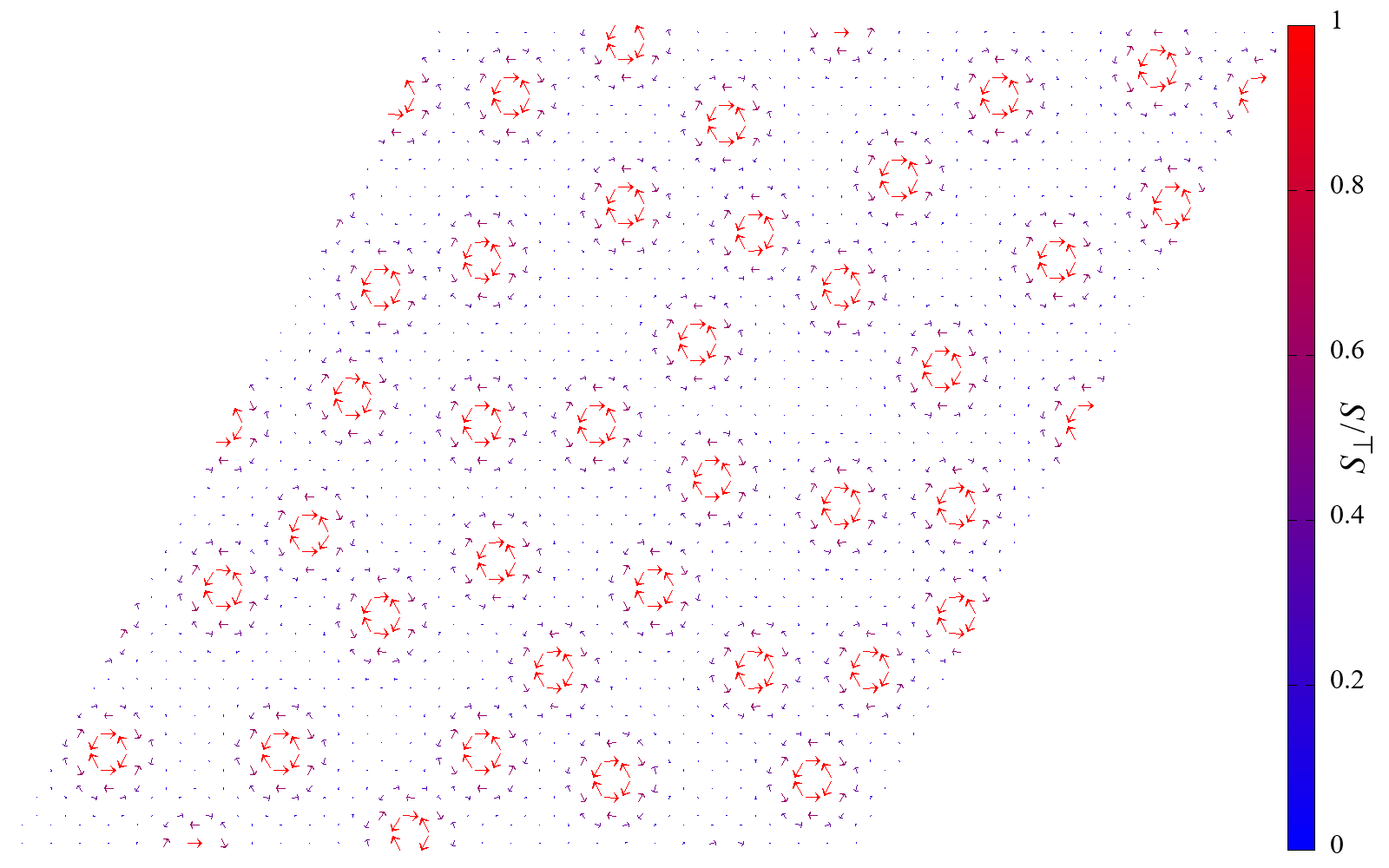}} \\
\subfloat[]{\label{phi025h097}
\includegraphics[scale=0.36]{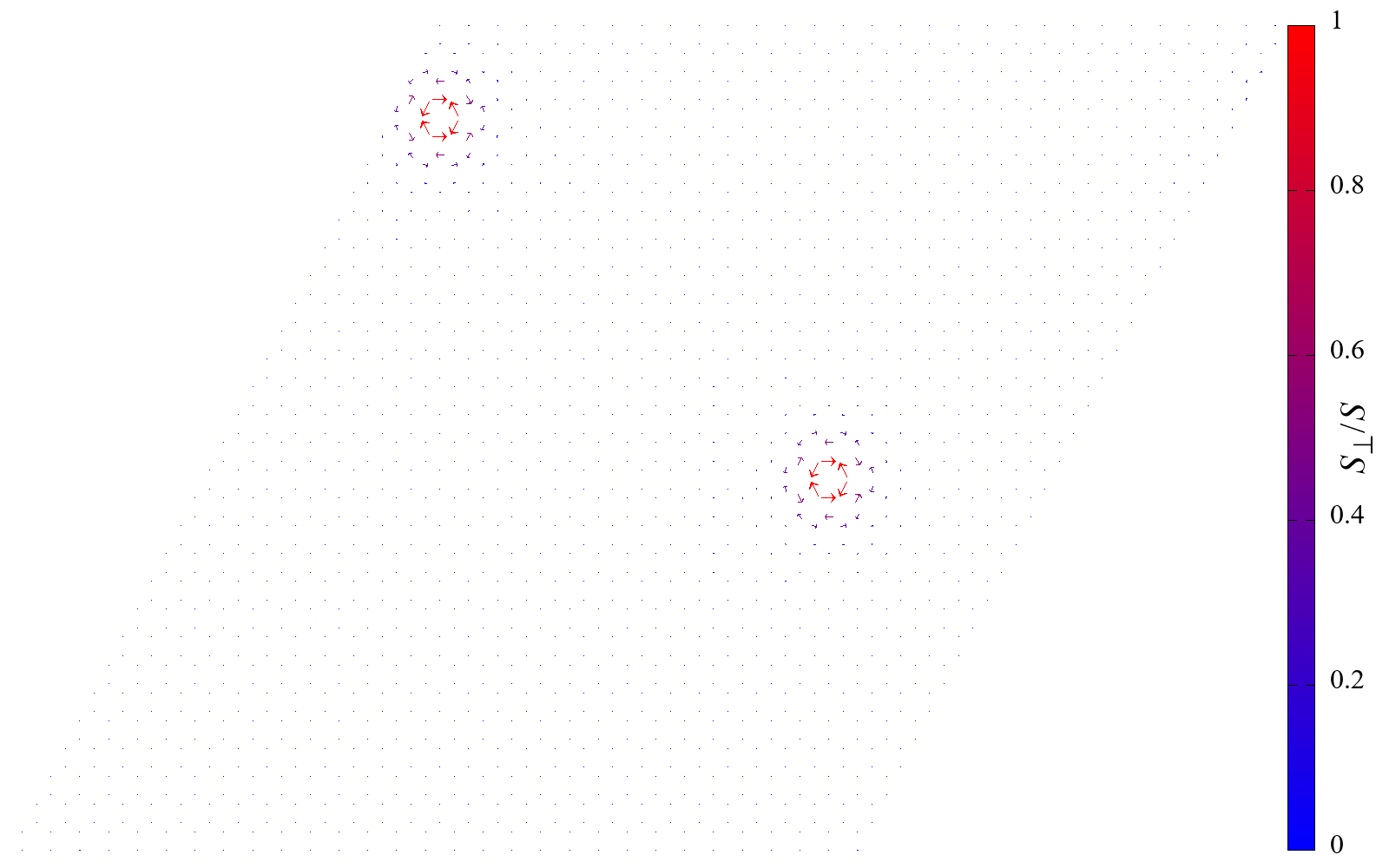}}
\caption{Sample spin configurations of the magnetic vortex liquids at $\varphi/\pi=0.25$, (a) $h/S=0.92$ and (b) $h/S=0.97$. For a fixed $\varphi$, the density of vortices decreases as the field increases.}
\end{figure}

\begin{figure}
\subfloat[]{\label{phi010h146}
\includegraphics[scale=0.36]{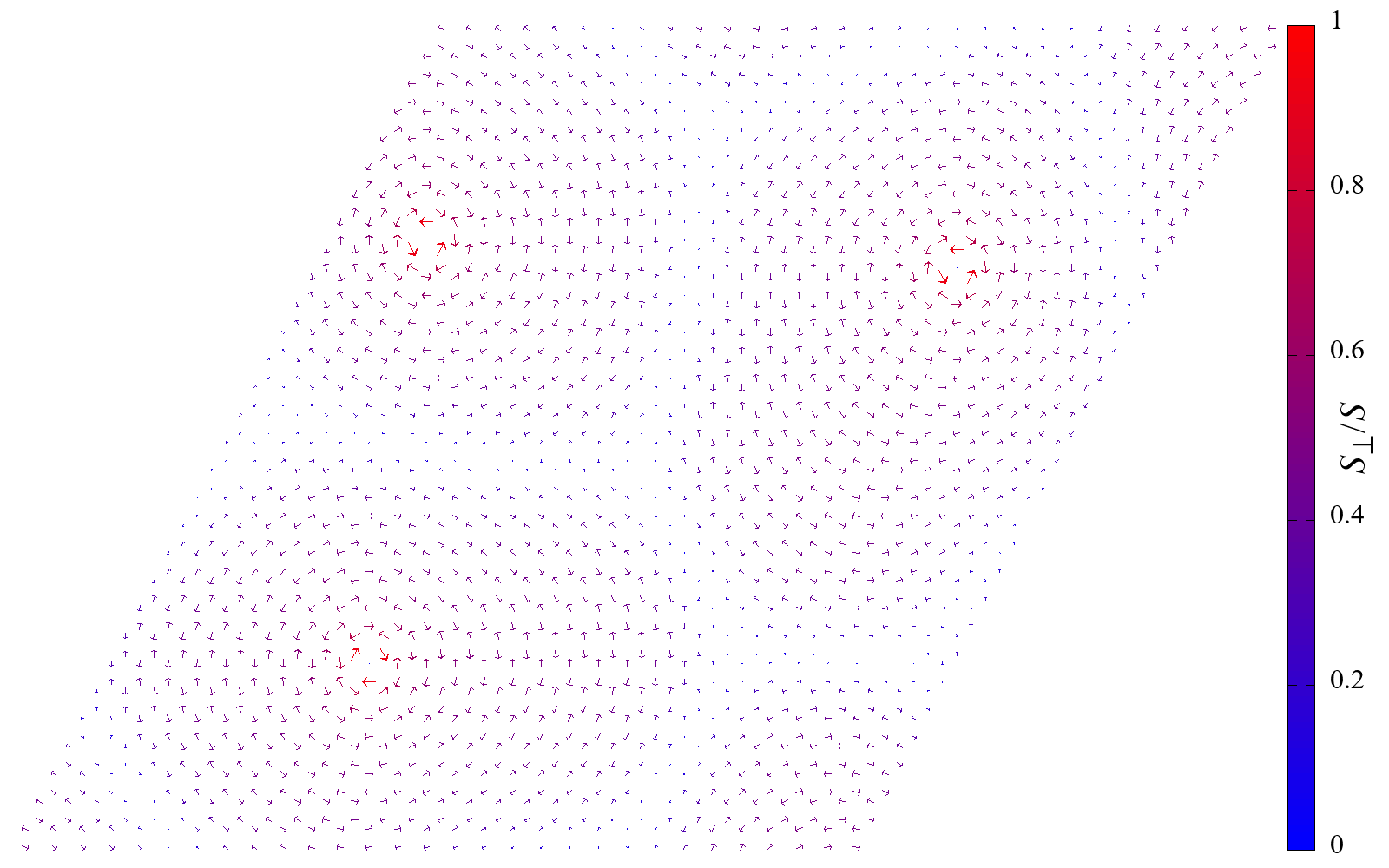}} \\
\subfloat[]{\label{phi040h040}
\includegraphics[scale=0.36]{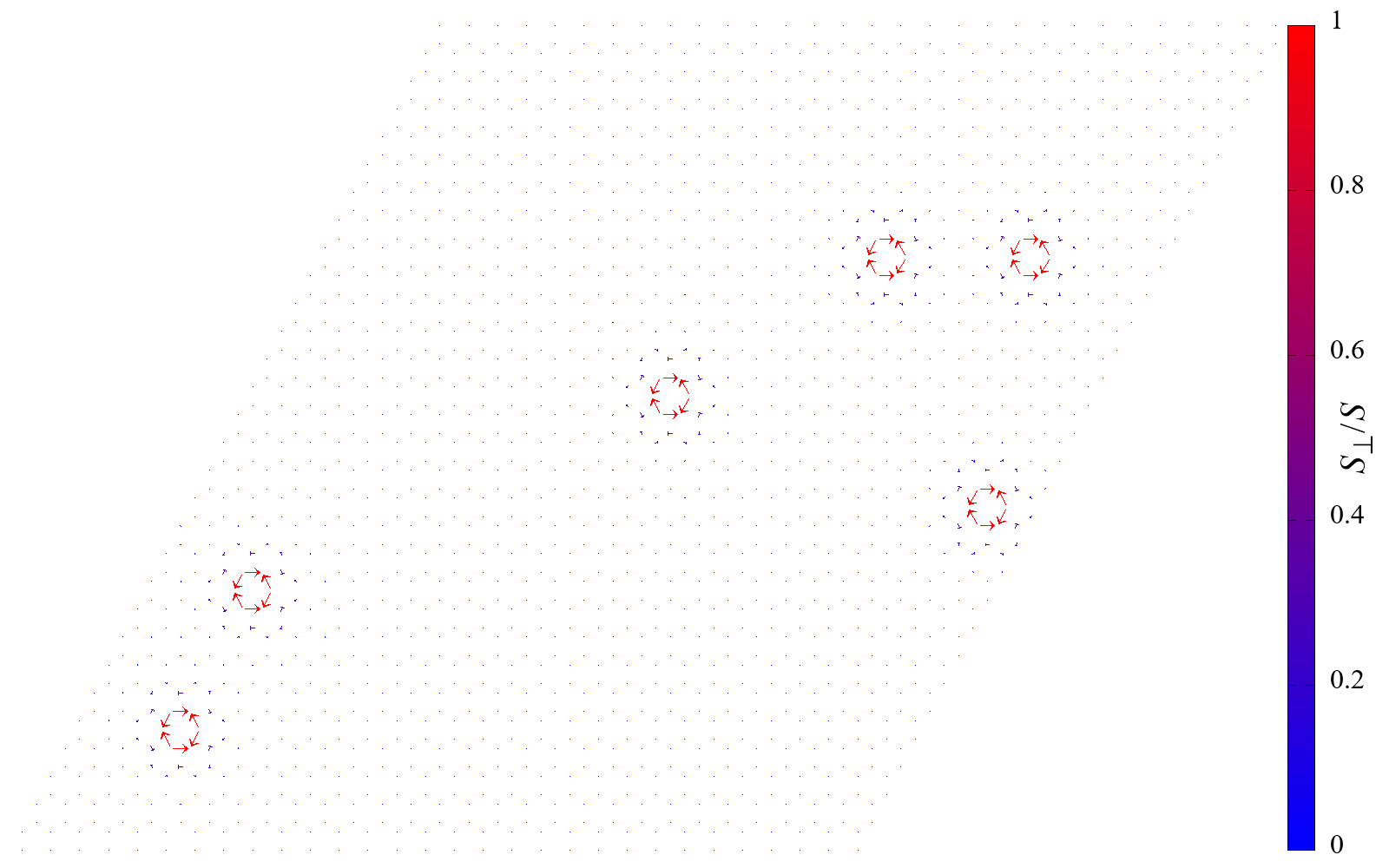}}
\caption{Sample spin configurations of the magnetic vortex liquids at (a) $\varphi/\pi=0.1$, $h/S=1.46$ and (b) $\varphi/\pi=0.4$, $h/S=0.4$. The size (structure) of vortex decreases (simplifies) as $\varphi$ increases. Moreover, the fields where the vortex liquid appears decrease as $\varphi$ increases.}
\end{figure}

\begin{figure}
\includegraphics[scale=0.36]{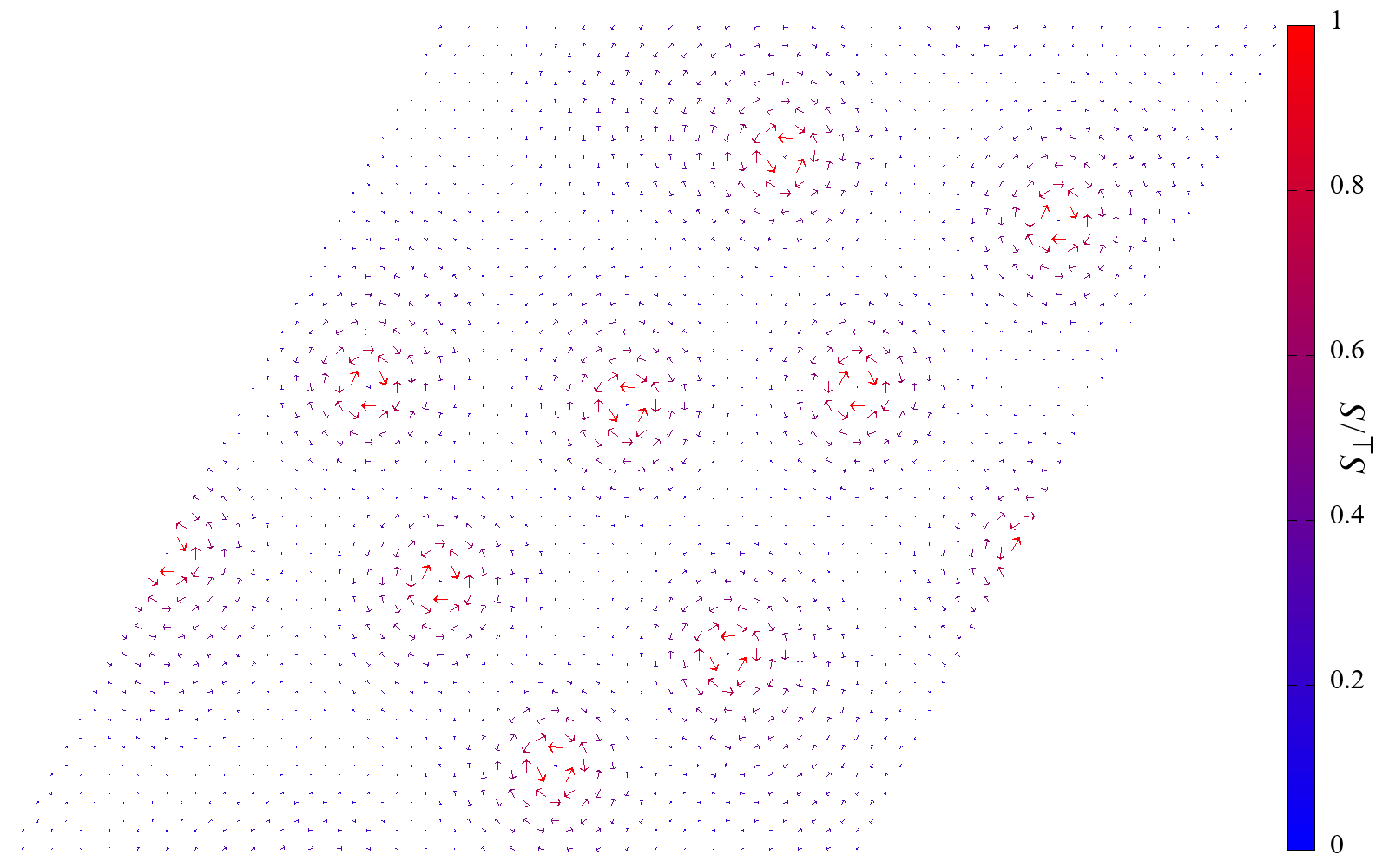}
\caption{\label{phi015h130}A sample spin configuration of the magnetic vortex liquid at $\varphi=\tan^{-1} (1/2)$ and $h/S=1.3$, to which we apply the mean field theory. It has 9 vortices, each of which carries a flux of $-4 \pi$.}
\end{figure}

\begin{figure}
\subfloat[]{\label{xi000h148}
\includegraphics[scale=0.36]{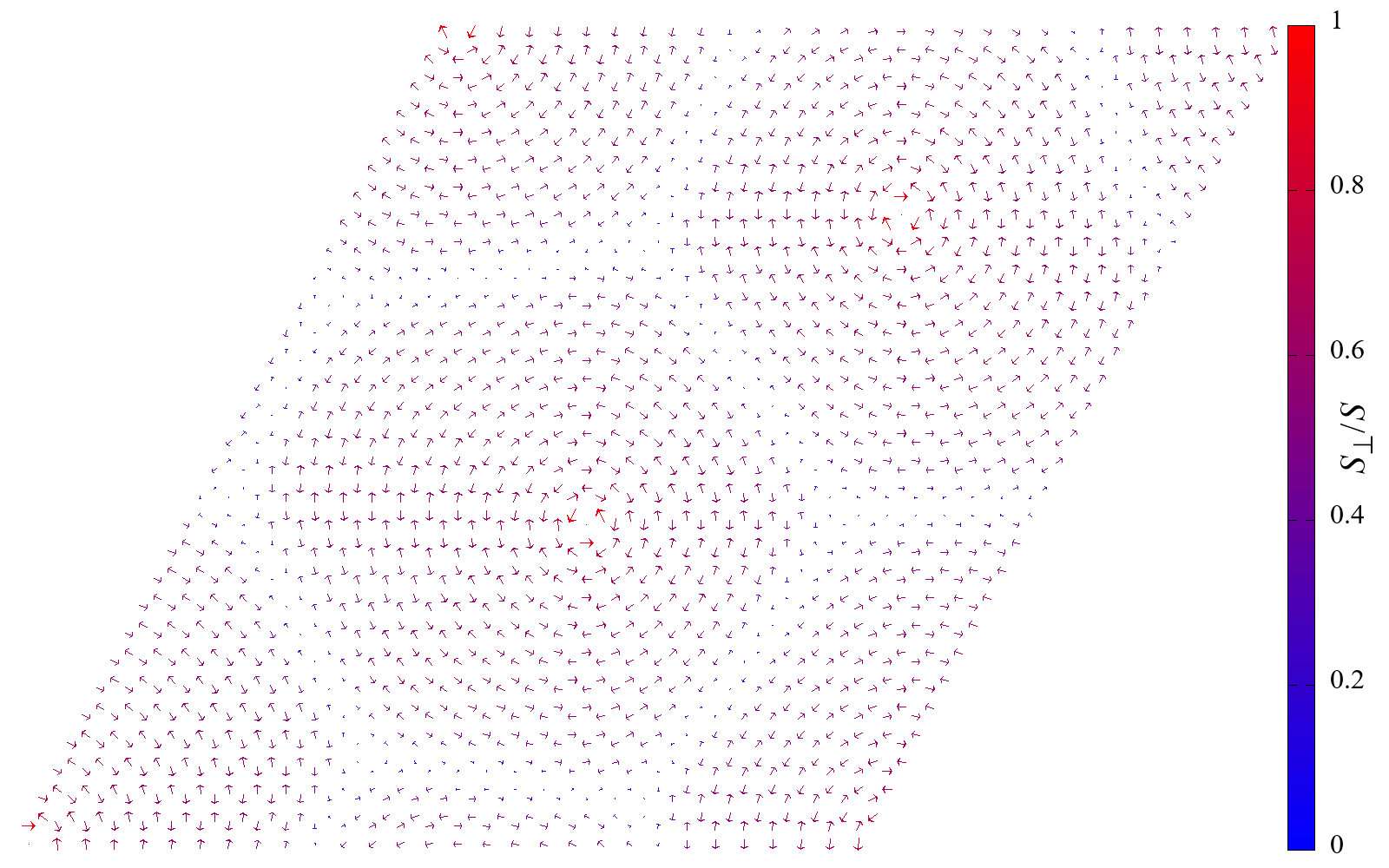}} \\
\subfloat[]{\label{xi010h140}
\includegraphics[scale=0.36]{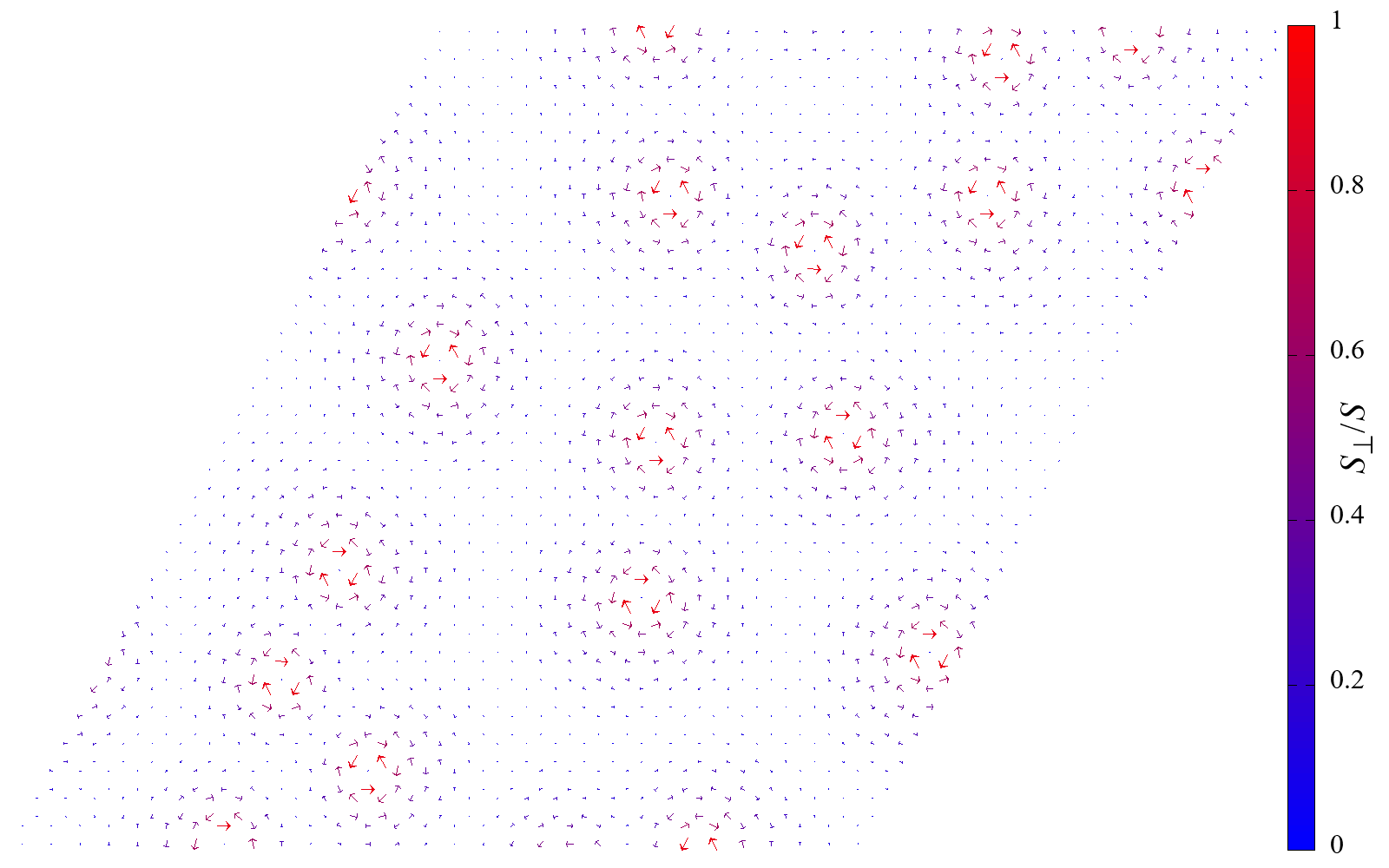}}
\caption{Sample spin configurations of the magnetic vortex liquids at (a) $\xi=0$, $h/S=1.48$ and (b) $\xi=0.1$, $h/S=1.4$. The size (structure) of vortex decreases (simplifies) as $\xi$ increases. However, the vortices are still quite large at $\xi=0.1$.}
\end{figure}

\begin{figure}
\subfloat[]{\label{xi030h124}
\includegraphics[scale=0.36]{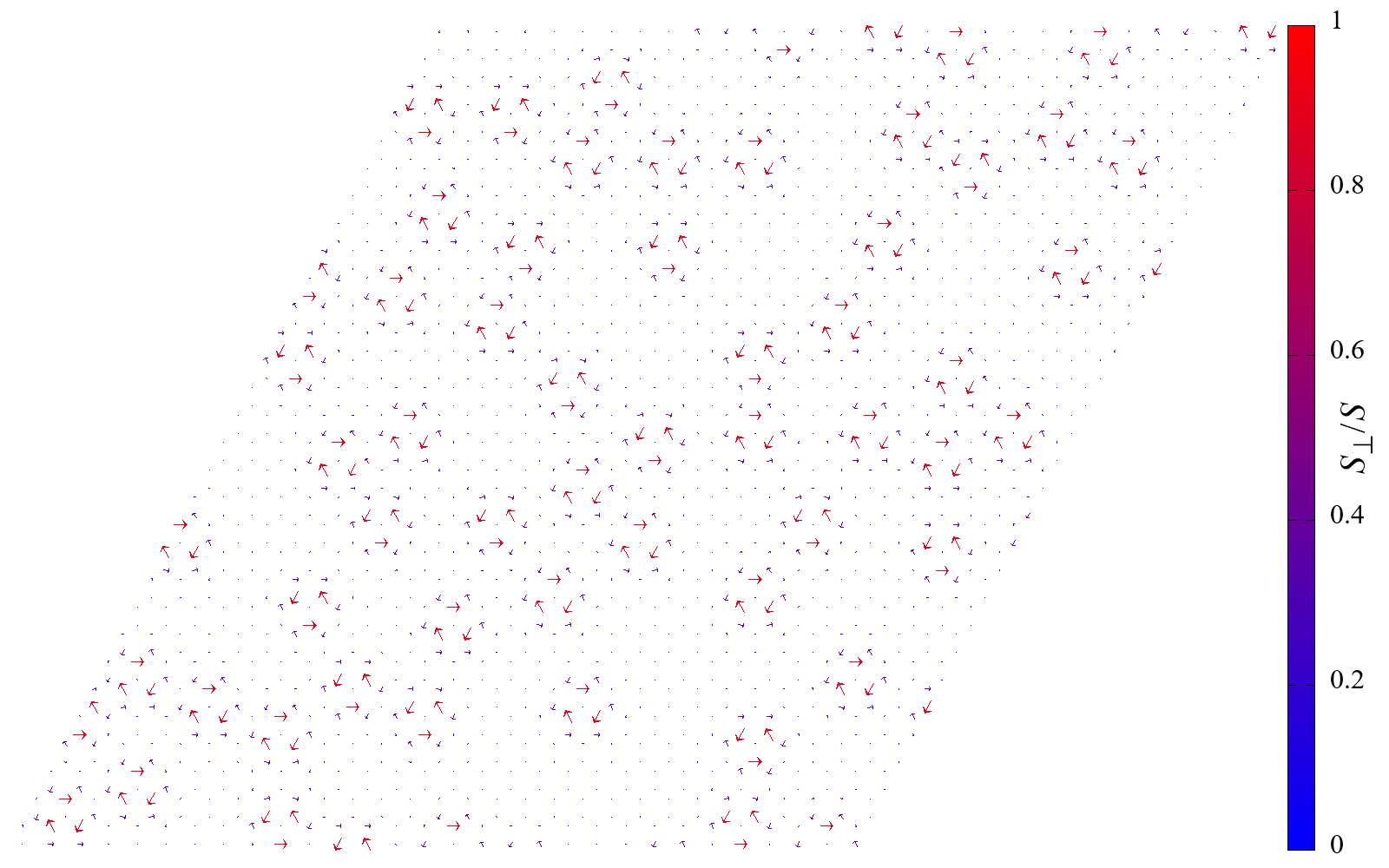}} \\
\subfloat[]{\label{xi070h112}
\includegraphics[scale=0.36]{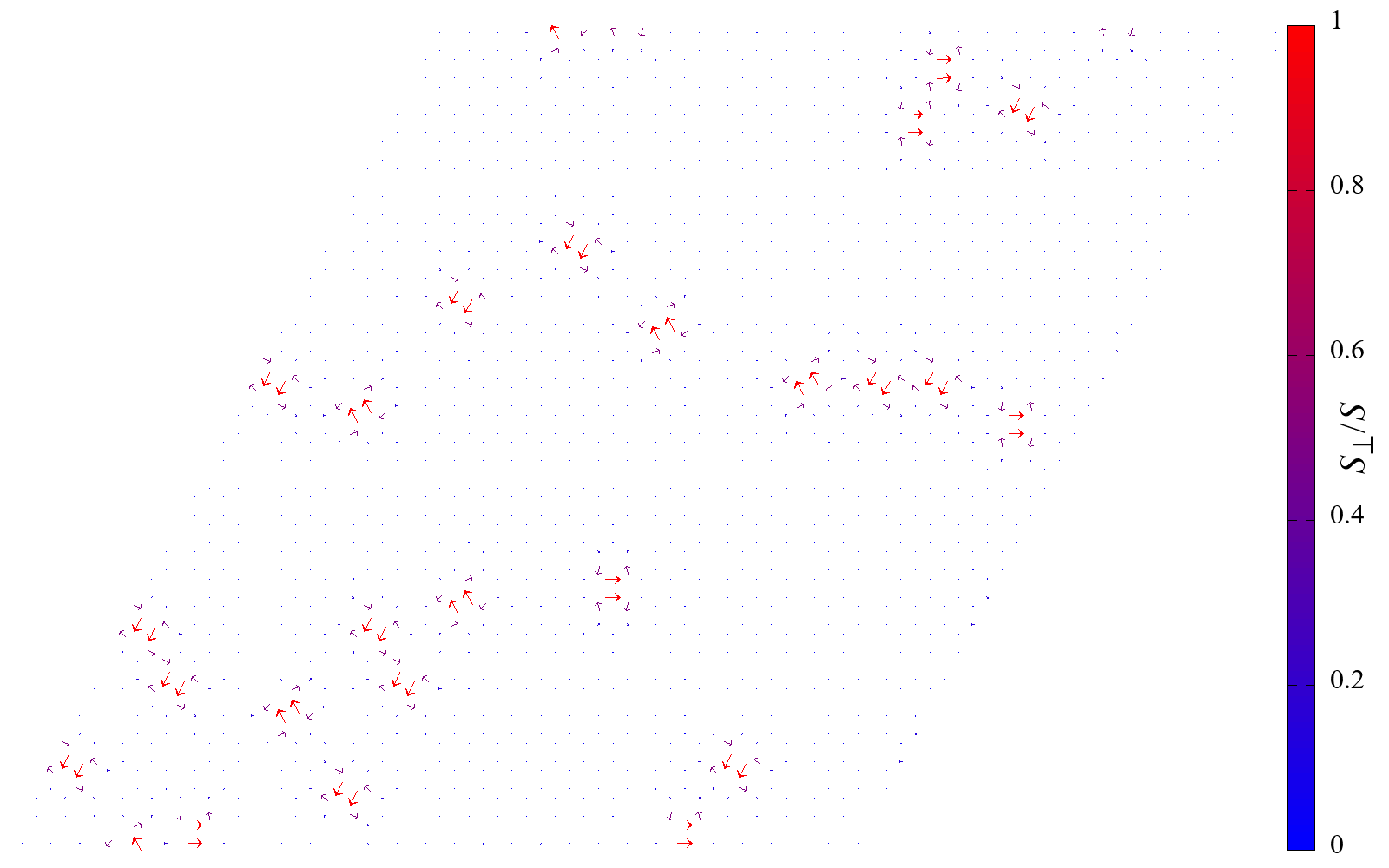}}
\caption{Sample spin configurations of the magnetic vortex liquids at (a) $\xi=0.3$, $h/S=1.24$ and (b) $\xi=0.7$, $h/S=1.12$. The size (structure) of vortex greatly reduces (simplifies) beyond $\xi \approx 0.2$, and remains more or less the same until $\xi \approx 0.65$, where a qualitative structural change occurs. The fluid-like nature of the vortex liquid is preserved throughout $\xi \in [0,0.8]$.}
\end{figure}

\section*{S2.~Vortex Crystallization}
In Sec.~III D, we mention that the magnetic vortex liquid may crystallize if the vortices become sufficiently dense.
If this is the case, the vortex lattice constant depends on the density, which is a continuous function of the field strength. Crystallization is thus in general expected to be incompatible with the periodic boundary conditions of the underlying lattice, and finite size effect becomes important.
We illustrate this in Fig.~S6 where cusps in the specific heat signal the onset of vortex crystallization. This is unlike the dilute vortex liquid which is associated with a smooth, featureless specific heat curve, as shown in Fig.~6. The occurrence of cusps depends on the system size, indicating that finite size effect is significant. 
Focusing on a fixed system size $L=24$, a cusp indicating crystallization is present at $h/S=0.85$ (see Fig.~S6a), whereas it is absent at $h/S=0.9$ (see Fig.~S6b). \\

We can illustrate this by plotting randomly drawn configurations from the Monte Carlo simulations at a temperature below the cusp, $T=0.02$. At $h/S=0.85$, the configuration indeed looks like a crystalline order (superimposed with small thermal fluctuations), see Fig.~S7a. At $h/S=0.9$, where the cusp is absent for this system size, the vortex arrangement is no longer regular (see Fig.~S7b).

\begin{figure}
\includegraphics[width=\linewidth]{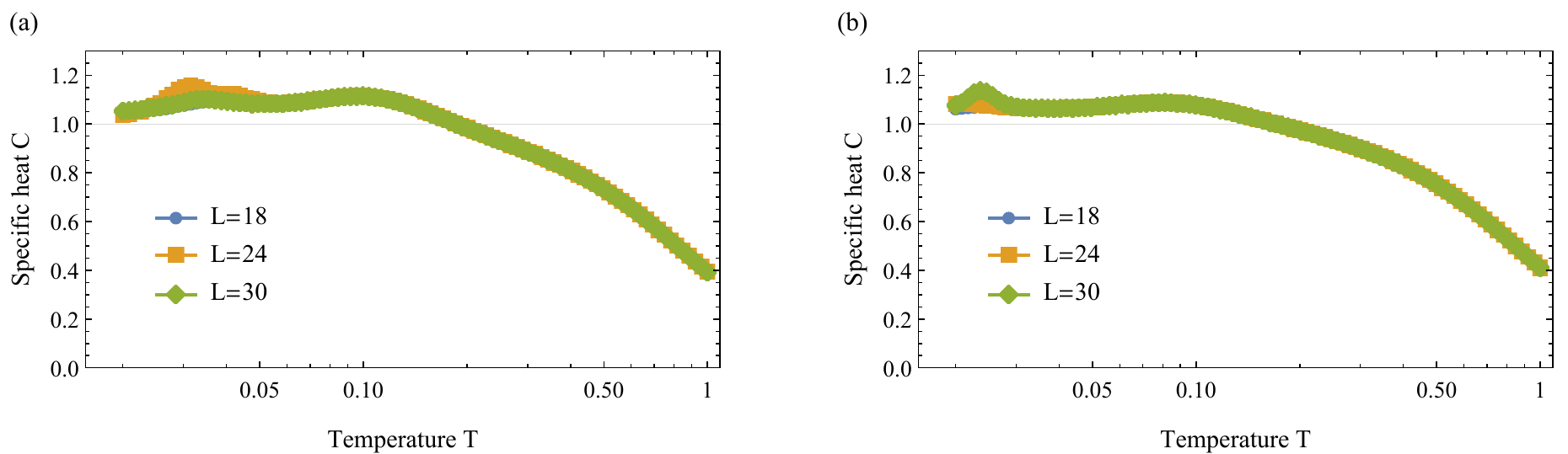}
\caption{Specific heat at finite temperatures. The data is obtained at $\varphi = \pi/4$ for different system sizes $L=18,24,30$ and magnetic field strength (a) $h/S=0.85$, (b) $h/S=0.90$. System size dependent cusps in the specific heat signal the onset of vortex crystallization with lattice constants that are incompatible with the system size.}
\label{fig:heat_B085_B090}
\end{figure}

\begin{figure}
\includegraphics[width=\linewidth]{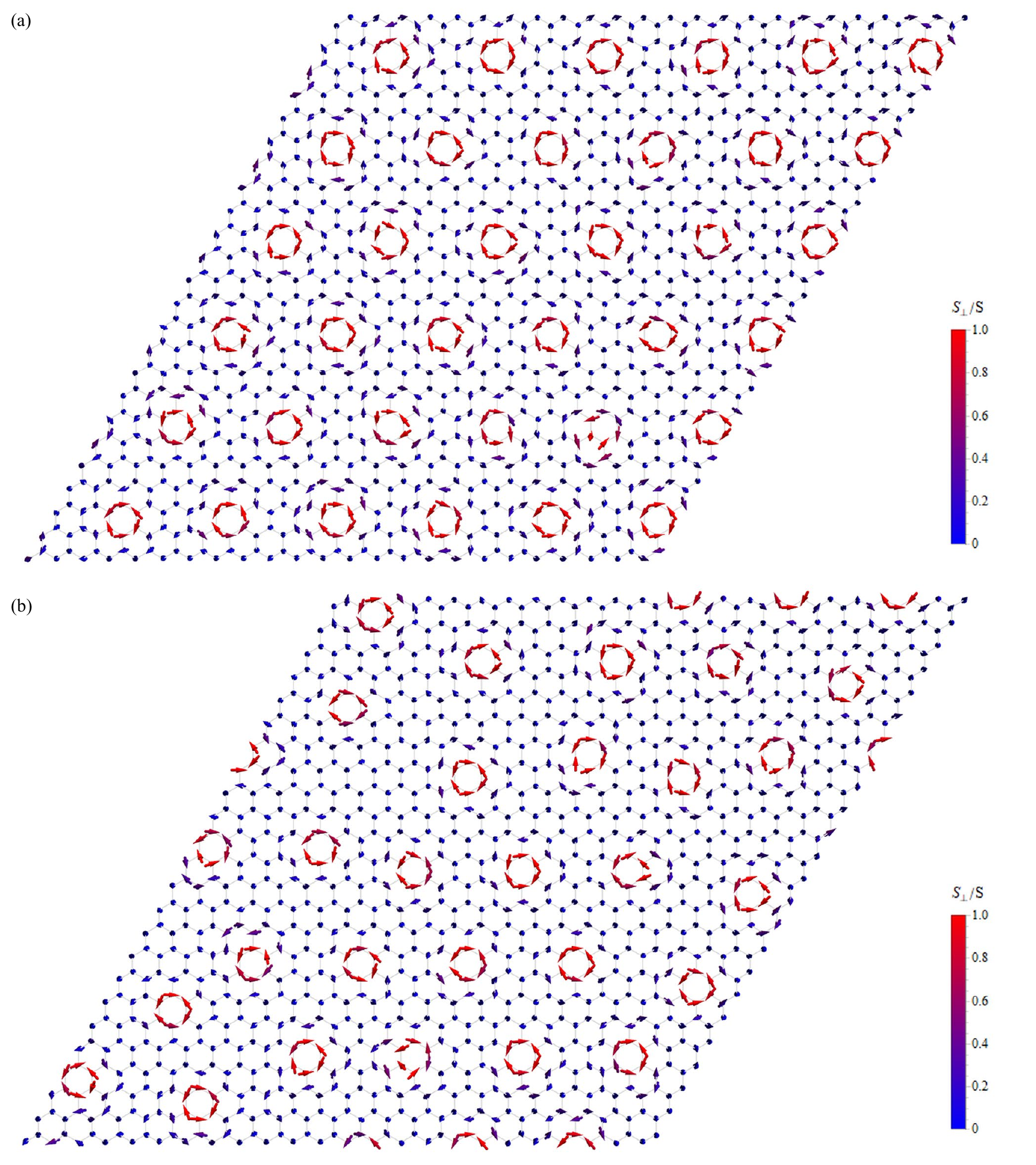}
\caption{Spin configurations drawn from Monte Carlo simulations at $\varphi = \pi/4$, temperature $T=0.02$, and system size $L=24$. (a) At $h/S=0.85$, the magnetic vortices have undergone crystallization and form a lattice structure. (b) At $h/S=0.9$, the the vortex arrangement is semi-regular, since the optimal vortex lattice constant is incompatible with the boundary conditions of the finite system.}
\label{fig:conf_B085_B090}
\end{figure}

\section*{S3.~Dynamical Spin Structure Factor from Mean Field Theory}
We show the dynamical spin structure factor $\mathcal{S} (\mathbf{k},\omega)$ of the magnetic vortex liquid calculated from the mean field theory in Fig.~S8, where the fluxes carried by the vortices are averaged over the entire system. As discussed in the main text, it bears features that are highly similar to the dynamical spin structure factor calculated from the classical Landau-Lifshitz equation. In Fig.~S8, we have removed the elastic contribution at zero energy at the $\Gamma$ point, which is a factor of $NS$ stronger than the signals from magnon excitations, so that the latter can be easily seen.

\begin{figure}
\includegraphics[scale=0.36]{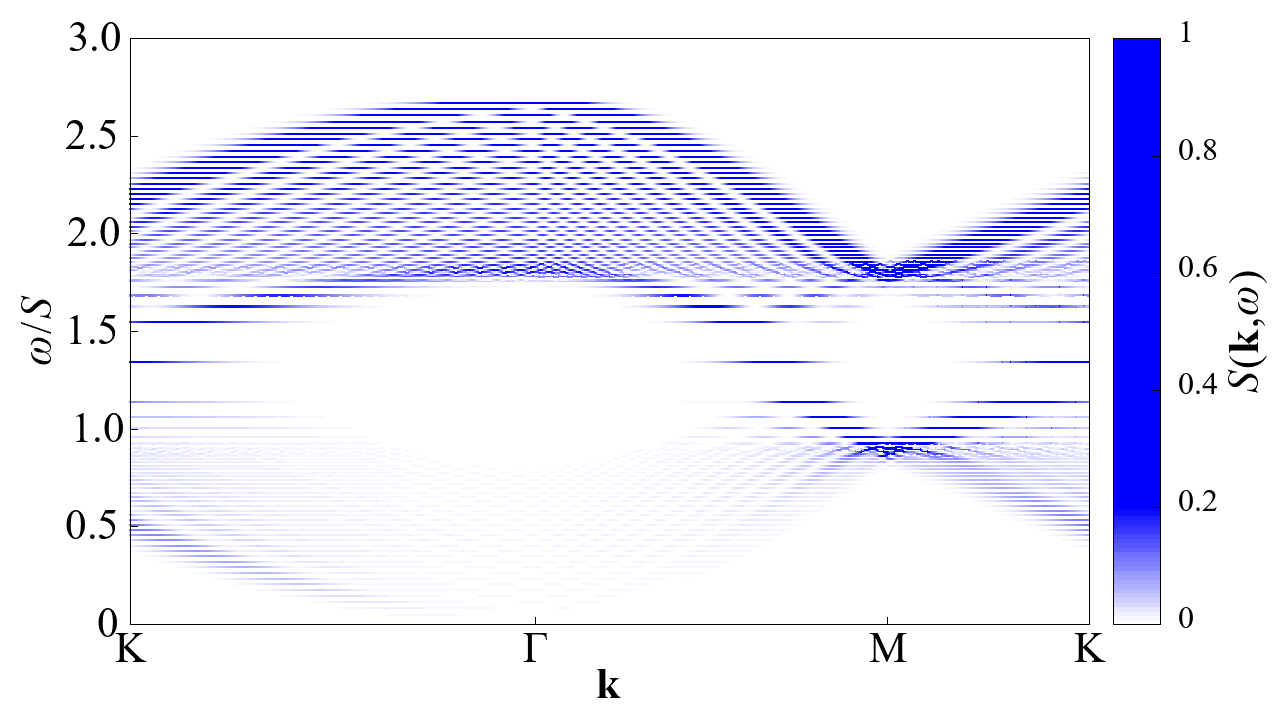}
\caption{\label{dynamicmeanfield}Dynamical spin structure factor of the magnetic vortex liquid at $\varphi=\tan^{-1} (1/2)$ and $h/S=1.3$, calculated from the mean field theory. The intensity is normalized such that its maximum is $1$.}
\end{figure}

\end{document}